\documentclass[twocolumn,prb,aps,showpacs,footinbib,amsmath,amssymb,superscriptaddress,longbibliography]{revtex4-1}
\usepackage{array}
\usepackage{graphicx}
\usepackage{dcolumn}
\usepackage{color}
\usepackage{bm}
\usepackage{natbib,hyperref}
\usepackage{amsmath,amssymb,amsfonts,bbold}
\usepackage[normalem]{ulem}

\newcommand{\beq}{\begin{equation}}
\newcommand{\eeq}{\end{equation}}
\newcommand{\beqa}{\begin{eqnarray}}
\newcommand{\eeqa}{\end{eqnarray}}

\begin{document}

\title{Structural and magnetic properties of 3d transition metal oxide chains \\
			on the (001) surfaces of Ir and Pt}

\author{Martin Schmitt}
	\address{Physikalisches Institut, Experimentelle Physik II, 
	Universit\"{a}t W\"{u}rzburg, Am Hubland, 97074 W\"{u}rzburg, Germany}
\author{Chong H.\ Park}
	\affiliation{Physikalisches Institut, Experimentelle Physik II, 
	Universit\"{a}t W\"{u}rzburg, Am Hubland, 97074 W\"{u}rzburg, Germany} 
	\affiliation{University of British Columbia, 2329 West Mall, Vancouver, BC 	Canada}
\author{Paula Weber}
	\affiliation{Physikalisches Institut, Experimentelle Physik II, 
	Universit\"{a}t W\"{u}rzburg, Am Hubland, 97074 W\"{u}rzburg, Germany} 
\author{Andreas~J{\"a}ger}
	\affiliation{Physikalisches Institut, Experimentelle Physik II, 
	Universit\"{a}t W\"{u}rzburg, Am Hubland, 97074 W\"{u}rzburg, Germany} 
\author{Jeannette Kemmer}
	\affiliation{Physikalisches Institut, Experimentelle Physik II, 
	Universit\"{a}t W\"{u}rzburg, Am Hubland, 97074 W\"{u}rzburg, Germany} 
\author{Matthias~Vogt} 
	\affiliation{Physikalisches Institut, Experimentelle Physik II, 
	Universit\"{a}t W\"{u}rzburg, Am Hubland, 97074 W\"{u}rzburg, Germany}
\author{Matthias Bode} 
\email[corresponding author: ]{bode@physik.uni-wuerzburg.de}
	\address{Physikalisches Institut, Experimentelle Physik II, 
	Universit\"{a}t W\"{u}rzburg, Am Hubland, 97074 W\"{u}rzburg, Germany}	
	\address{Wilhelm Conrad R{\"o}ntgen-Center for Complex Material Systems (RCCM), 
	Universit\"{a}t W\"{u}rzburg, Am Hubland, 97074 W\"{u}rzburg, Germany}    
	   


\date{\today}

\begin{abstract}
We present a survey of the structural and magnetic properties of submonolayer transition metal dioxides 
on the (001) surfaces of the heavy face-centered cubic (fcc) noble metals Ir and Pt performed 
by spin-averaged scanning tunneling microscopy (STM) and spin-polarized (SP-)STM.  
Our STM results confirm that deposition of Co, Fe, Mn, and Cr on the $(2\,\times\,1)$ oxygen-reconstructed Ir(001) surface 
leads to the formation of quasi one-dimensional chains with a $(3\,\times\,1)$ unit cell.  
As recently predicted by density functional theory [Ferstl \textit{et al.}, Phys.\ Rev.\ Lett.\ \textbf{117}, 046101 (2016)], 
our SP-STM images of FeO$_2$ and MnO$_2$ on Ir(001) show a two-fold periodicity along the chains 
which is characteristic for an antiferromagnetic coupling along the chains.  
In addition, these two materials also exhibit spontaneous, permanent, and long-range magnetic coupling across the chains. 
Whereas we find a ferromagnetic inter-chain coupling for FeO$_2$/Ir(001), the magnetic coupling of MnO$_2$ on Ir(001) 
appears to be a non-collinear $120^\circ$ spin spiral, resulting in a $(9\,\times\,2)$ magnetic unit cell. 
On Pt(001) patches of $(3\,\times\,1)$-reconstructed oxide chains could only be prepared 
by transition metal (Co, Fe, and Mn) deposition onto the cold substrate and subsequent annealing in an oxygen atmosphere.  
Again SP-STM on MnO$_2$/Pt(001) reveals a very large, $(15\,\times\,2)$ magnetic unit cell 
which can tentatively be explained by a commensurate $72^\circ$ spin spiral.  
Large scale SP-STM images reveal a long wavelength spin rotation along the MnO$_2$ chain. 
\end{abstract}


\maketitle

\section{Introduction}
\vspace{-0.3cm}
Significant progress has been achieved towards a thorough understanding 
of magnetically ordered states in solid-state materials.\cite{Vaz2008}   
Over the past 40 years spin structures with increasing complexity were detected. 
Whereas collinear ferro- or antiferromagnetism governed by the competition 
of exchange, magnetocrystalline anisotropy, and dipolar interactions 
initially dominated the scientific debate, we have witnessed a focussing 
on more complex non-collinear magnetic states since the advent of the current century.\cite{Braun2012} 
This development was---at least partially---made possible by the development 
of advanced surface analysis and microscopy tools which allow for the detection 
of magnetic signals with unprecedented sensitivity and spatial resolution. 
In the context of this contribution spin-polarized scanning tunneling microscopy (SP-STM) will be of particular interest.
This technique utilizes the tunnel magnetoresistance effect between a magnetic surface and a spin-polarized tip 
to obtain information about the sample's spin structure with atomic resolution.\cite{Bode2003} 
SP-STM allowed for the first direct imaging of antiferromagnetic surfaces\cite{KFB2005} and domain walls,\cite{BVB2006} 
as well as of frustrated N{\'{e}}el spin states with antiphase domains.\cite{GWK2008} 
Furthermore, it turned out that the spin-orbit--induced Dzyaloshinskii-Morija interaction (DMI), 
which has previously considered in some rare cases only, can be very significant at surfaces and interfaces. 
For example, it turned out that the Mn monolayer on W(110), which has initially been assumed 
to be a simple collinear and uniaxial antiferromagnet,\cite{PhysRevB.66.014425,Heinze2000} 
instead forms chiral spin cycloid.\cite{Bode2007}

\begin{figure}[b]
	\includegraphics[width=0.99\columnwidth]{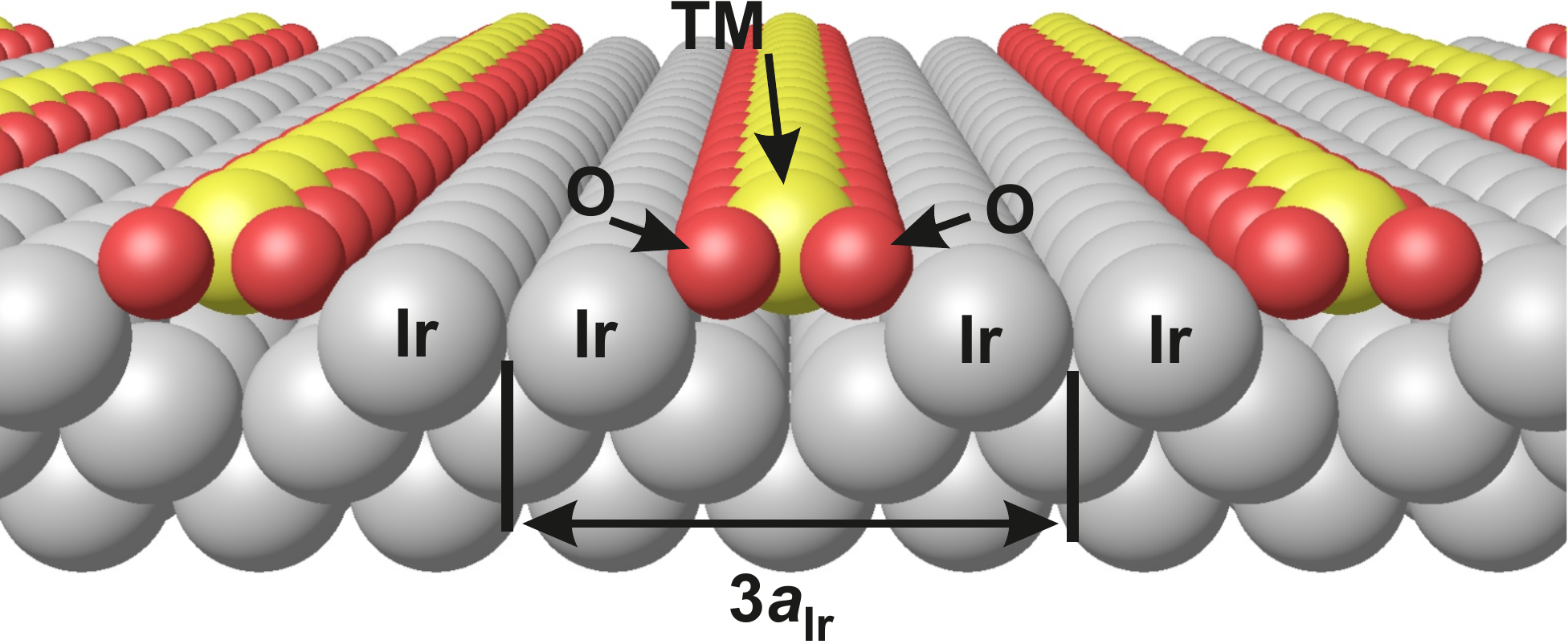}
			\caption{Structure model of transition metal oxide chains on Ir(001) as proposed in Ref.\,\onlinecite{Ferstl2016}. 
			 The transition metal atoms chains sit above empty substrate rows, held in place by the oxygens atoms.  
			 The inter-chain spacing corresponds to 3$a_{\rm Ir}$.}
	\label{Fig:Scheme}
\end{figure}

Recently a group of novel quasi one-dimensional $3d$ transition metal oxides (TMO) was discovered 
which can conveniently be prepared by self-organized growth 
on the (001) surfaces of the heavy fcc metals Ir and Pt.\cite{Ferstl2016,Ferstl2017} 
For Ni, Co, Fe, and Mn on Ir(001) and also for Co on Pt(001) a structural $(3\,\times\,1)$ unit cell 
was observed by low-energy electron diffraction (LEED).  
In either case scanning tunneling microscopy (STM) reveals a surface morphology which is characterized by long 
and highly periodic chains oriented along the $[100]$ and $[010]$ high symmetry directions of the (001) surface. 
The structure of the TMO chains on fcc(001) surfaces as proposed by Ferstl {\em et al.}\cite{Ferstl2016} 
is schematically represented in Fig.\,\ref{Fig:Scheme}.  
Within each chain we find two oxygen atoms (red) between nearest-neighbor transition metal atoms (yellow). 
The TMO chains sit above empty substrate rows, held in place by the oxygens atoms.

Indeed, density functional theory (DFT) calculations reproduced 
the experimentally determined structural properties well.\cite{Ferstl2016} 
These theoretical investigations also predicted highly interesting intra-chain magnetic couplings, 
ranging from non-magnetic (NM) NiO$_2$ via ferromagnetic (FM) coupling for CoO$_2$ and FeO$_2$ 
to an antiferromagnetic (AFM) interaction along MnO$_2$ chains on Ir(001).\cite{Ferstl2016} 
Furthermore, an AFM coupling was predicted for CoO$_2$ chains on Pt(001).\cite{Ferstl2017} 
In general, the coupling strength was found to be much stronger along the chains (up to 44\,meV per TM atom) 
than across adjacent chains (a few meV).  
It should be kept in mind, however, that the calculations 
performed in Refs.\,\onlinecite{Ferstl2016} and \onlinecite{Ferstl2017}, 
which are qualitatively summarized in Table\,\ref{Tab:prediction}, were restricted to collinear spin configurations.  
Non-collinear magnetic structures, such as the N{\'{e}}el state, spin spirals, skyrmions, or helical spins structures,
which can potentially arise from frustration,\cite{GWK2008} higher-order exchange,\cite{KSH2018,RPK2018} 
or the DMI\cite{Bode2007} have not been considered.

\begin{table}[tb]
\begin{tabular}{|p{2.0cm}||c|c|c|c|c|}
\hline
substrate &  \multicolumn{4}{|c|}{Ir(001)}  & Pt(001)\\
\hline
TMO chain & NiO$_2$ & CoO$_2$ & FeO$_2$ & MnO$_2$ & CoO$_2$ \\
\hline\hline
$\mu$\,$(\mu_{\rm B})$ & 0.00 & 1.96 & 3.55 & 3.62 & -- \\
\hline
intra-chain coupling ($\parallel$) & NM & FM & AFM & AFM & AFM\\
\hline
$\Delta E_\parallel$ (meV)& -- & 25 & 44 & 27 & -- \\
\hline
inter-chain coupling ($\perp$)& -- & FM & AFM & AFM & -- \\
\hline
$\Delta E_\perp$ (meV)& -- & 4 & 9 & 0.4 & -- \\
\hline
\end{tabular}
\caption{Magnetic moments and magnetic coupling of one-dimensional TMOs along (intra-) and across (inter-) adjacent chains 
		and corresponding energy gain as predicted by DFT in Refs.\,\onlinecite{Ferstl2016} and \onlinecite{Ferstl2017}.}
\label{Tab:prediction}
\end{table} 
To verify the predictions of Ferstl {\em et al.}\cite{Ferstl2016,Ferstl2017} 
we recently studied the magnetic structure of MnO$_2$ chains on Ir(001) 
by means of spin-polarized scanning tunneling microscopy (SP-STM).\cite{NC}
In addition to the AFM coupling along the chains predicted by Ferstl {\em et al.}\cite{Ferstl2016}
an indirect $120^{\circ}$ magnetic coupling across the chains was observed. 
This surprising finding was rationalized in terms of an Dzyaloshinskii-Moriya--enhanced 
Ruderman-Kittel-Kasuya-Yosida (RKKY) interaction.\cite{NC} 
These earlier results obtained on MnO$_2$/Ir(001) showed that this indirect magnetic coupling mechanism 
which was previously only observed for assemblies of single atoms or clusters on Pt(111) surfaces\cite{Khajetoorians2016,Hermenau2017} 
can also result in chiral magnetic order in laterally extended structures.

The purpose of this paper is to investigate the magnetic structure of a broad range of $(3\,\times\,1)$-ordered TMO chains 
on (001) surfaces of the heavy face-centered cubic (fcc) noble metals Ir and Pt by SP-STM. 
The paper is organized as follows:
The SP-STM technique and the experimental procedures applied for substrate cleaning, oxidation, 
and transition metal deposition are briefly described in Sect.\,\ref{sect:ExpProc}. 
Results for the two substrates, i.e., Ir(001) and Pt(001), will be presented separately
in Sect.\,\ref{sect:ResTMO_Ir} and Sect.\,\ref{sect:ResTMO_Pt}, respectively.
SP-STM measurements were performed on the oxides of Co, Fe, Mn, and Cr on Ir(001) and for Co and Mn on Pt(001).   
Whereas no magnetic contrast could be detected for Co and Cr, the magnetic intra-chain coupling 
observed for the other transition metals is in agreement with DFT predictions.\cite{Ferstl2016,Ferstl2017} 
In addition, our results also reveal magnetic ordering across the chains. 
Whereas we find a collinear coupling across the chains for FeO$_2$ on Ir(001), 
the indirect inter-chain coupling of MnO$_2$ on both, Ir(001) and Pt(001), 
is found to be helical, resulting in complex spin structures with surprisingly large magnetic unit cells.

\section{Experimental procedures}
\label{sect:ExpProc}
STM experiments were performed in a two-chamber ultra-high vacuum (UHV) system 
with a base pressure $p \leq 5 \times 10^{-11}$\,mbar. 
Clean Ir(001) and Pt(001) surfaces were prepared by annealing cycles in an oxygen atmosphere 
followed by cycles of sputtering and annealing without oxygen. 
After this procedure the well known $(5\,\times\,1)$ reconstruction of Ir(001) 
as well as the $(26\,\times\,118)$ structural unit cell of Pt(001) 
was confirmed.\cite{Grant1969,Rhodin1976,Schmidt2002,Hammer2016}

Closely following the procedures described by Ferstl and co-workers,\cite{Ferstl2016} the clean Ir(001) surface 
was then exposed to molecular oxygen resulting in a $(2\,\times\,1)$-reconstructed surface.  
The oxygen pressure indicated by our quadrupole mass spectrometer 
was $p_{{\rm O}_2} = 1 \times 10^{-8}$\,mbar, but the local pressure is assumed to be 
about two orders of magnitude higher since the gas nozzle is located just a few cm above the sample surface. 
On this oxygen-reconstructed Ir surface we deposited one third of a monolayer (ML) of either, Co, Fe, Mn, or Cr.  
All 3$d$ transition metals were vaporized with commercial e-beam evaporation sources (EFM3). 
Whereas Co and Fe were evaporated from wires with a diameter of 2\,mm, 
Mn and Cr lumps were loaded into Mo crucibles. 
Upon $3d$ transition metal deposition the sample was again annealed in an oxygen atmosphere,
resulting in the $(3\,\times\,1)$ structure of TMO chains.\cite{Ferstl2016} 
Since the annealing temperature $T_{\rm ann}$ required for optimal TMO chain quality 
was found to depend on the transition metal element, the specific values will be given below.
Due to the higher stability of the Pt(001) reconstruction to oxygen exposure 
the preparation of TMO chains on Pt(001) is slightly different.\cite{Morgan1968,Ferstl2017} 
Namely, the $3d$ transition metal was directly evaporated onto the reconstructed surface 
and only subsequently annealed in an oxygen atmosphere.\cite{Ferstl2017}

\begin{figure*}[t]
	\includegraphics[width=0.99\textwidth]{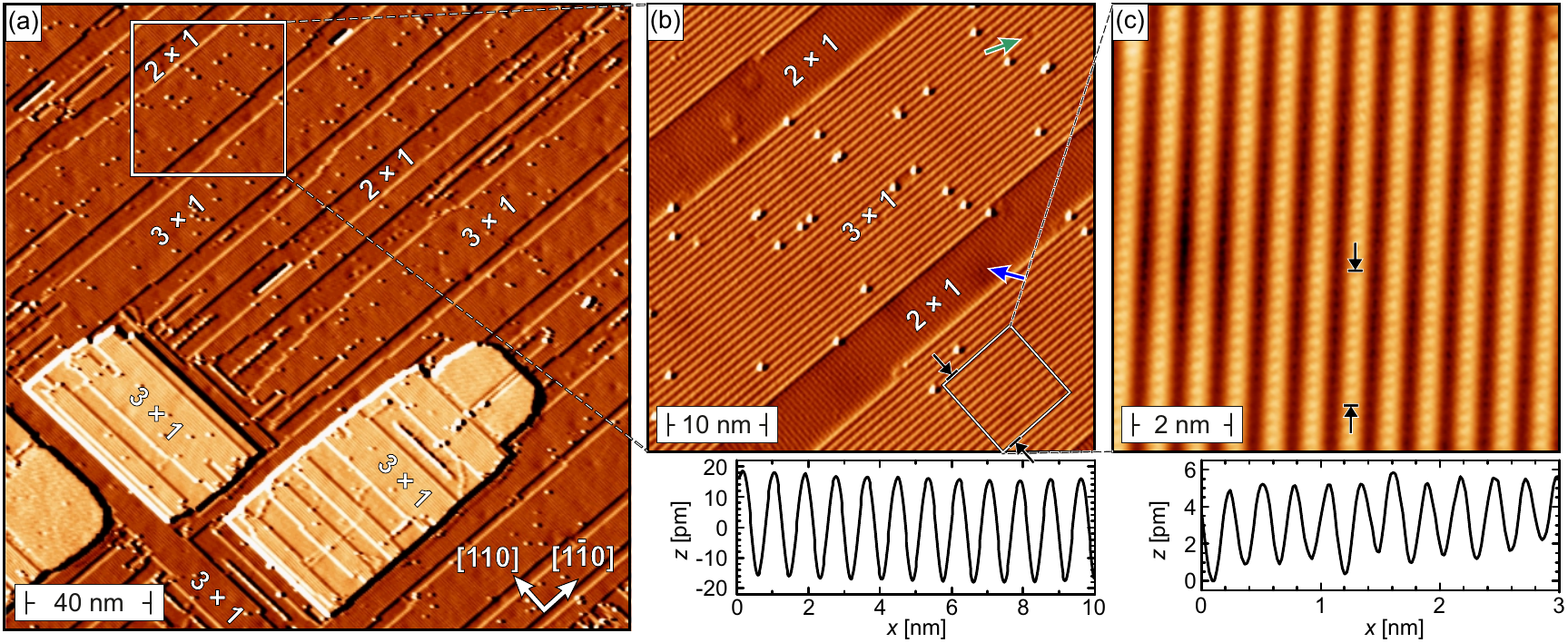}%
		\caption{(a) Overview scan of a sample with coexisting $(3\,\times\,1)$-ordered CoO$_2$ chains 
				and $(2\,\times\,1)$ oxygen-reconstructed areas on Ir(001). 
				(b) Higher magnification image of the area marked by a white square in (a) 
				revealing an orthogonal orientation row orientation of $(2\,\times\,1)$- and $(3\,\times\,1)$-ordered areas. 
				The line profile in the bottom panel confirms a $3a_\textrm{Ir}$ periodicity 
				perpendicular to the CoO$_2$ chains. 
				(c) Atomic resolution scan of the area marked by a box in (b).  
				The line profile verifies the $\times\,1$ periodicity along the CoO$_2$ chains.
				Scan parameters: (a), (b) $U = 1$\,V, $I = 300$\,pA; (c) $U = 50$\,mV, $I = 1$\,nA. 
				All scans performed with a non-magnetic W tip.}
	\label{Fig:CoO-Ir}
\end{figure*}
To verify the structural properties of the TMO chains the samples were transferred 
into a home-built low-temperature scanning tunneling microscope (LT-STM) where they were scanned 
with an electro-chemically etched polycrystalline W tip at an operation temperature of $T \approx 5.5$\,K.  
All images were obtained in the constant-current mode with bias voltage ($U$) applied to the sample. 
When using spin-polarized tips in SP-STM measurements the recorded tunneling current $I$ can be written as
\begin{equation}
I = I_\textrm{0} + P_{\rm t} \cdot P_{\rm s} \cdot \cos \theta, 
\label{Eq:SP-STM}
\end{equation}  
where $I_\textrm{0}$ is the spin-averaged contribution to the tunneling current. 
The second term in Eq.\,\ref{Eq:SP-STM} represents the magnetization direction-dependent variation 
of the total current which depends on the angle $\theta$ between tip $P_{\rm t}$ and sample $P_{\rm s}$ polarization.     
For SP-STM the W tips were flashed by electron bombardment and coated with either Fe or Cr. 
To unambiguously determine the in-plane or out-of-plane sensitivity 
of the magnetic SP-STM tips they were characterized on a reference system. 
In the present case we used the Fe double-layer (DL) on W(110) which exhibits a well-known 
inhomogeneous spin spiral, thereby offering the possibility to identify the in-plane and out-of-plane components 
of the tip magnetization in a single measurement.\cite{Pietzsch2000,Bode2001,Meckler2009} 

As we will discuss below, we could not obtain magnetic contrast on some TMOs 
even though earlier DFT calculations predict them to order magnetically. 
This raises the question of the detection limit of SP-STM.  
In fact, the surface spin structures of numerous elements have successfully been imaged in the past, 
including rare-earth metals, such as Gd\cite{PhysRevLett.81.4256} 
and Dy,\cite{PhysRevB.76.064411} or the antiferromagnetic $3d$ metal Cr.\cite{PhysRevB.67.174411}
The Gd magnetic moment is largely carried by the $4f$ shell (7$\mu_{\rm B}$) 
which is energetically located well below the Fermi level and therefore cannot contribute to the tunneling current.  
For Gd(0001) it has been shown that the SP-STM contrast originates from a $d_{z^2}$-like surface state 
which carries a relatively low magnetic moment $\mu \approx 0.35 \mu_{\rm B}$ only.\cite{Kurz_2002} 
In the case of Cr measurements were performed at room temperature, i.e.\ at a relatively high 
reduced temperature $T/T_{\rm N} \approx 0.94$ (N{\'e}el temperature $T_{\rm N} = 311$\,K). 
At this temperature the magnetic moment only amounts to about 40\% of its ground state value.
Nevertheless, for both Gd(0001) and for Cr(001) the surface magnetic structure could clearly be imaged.
Considering these earlier results we assume that the detection limit of SP-STM 
is well below a surface moment of $1 \mu_{\rm B}$. 

\section{Results}
\subsection{TMO chains on Ir(001)}
\label{sect:ResTMO_Ir}
\subsubsection{CoO$_{2}$/Ir(001)}
\label{Sect:CoO2-Ir}

After evaporation of Co onto the oxidized Ir(001) surface at room temperature the sample was annealed 
at $T_{\rm ann} \approx 870$\,K at an oxygen partial pressure $p_{\rm{O_{2}}} = 1 \times 10^{-8}$\,mbar. 
During the TMO chain growth process every third Ir surface atom is expelled from the surface layer. 
It has previously observed for MnO$_2$ on Ir(001)\cite{Ferstl2016}\cite{NC} that---at sufficiently high annealing 
temperature---these atoms form extended islands or even recombine at step edges with existing surface terraces. 
A similar behavior can be observed in Fig.\,\ref{Fig:CoO-Ir}(a), where monatomic rectangular shaped islands occur 
with step edges along the $[110]$ and $[1\overline{1}0]$ high-symmetry directions of the substrate.

Closer inspection of the terraces in Fig.\,\ref{Fig:CoO-Ir}(b) show the coexistence 
of the $(2\,\times\,1)$ oxygen reconstruction and the $(3\,\times\,1)$ TMO structure along the high-symmetry directions. 
Furthermore, Fig.\,\ref{Fig:CoO-Ir}(b) reveals that the stripes of the $(2\,\times\,1)$ reconstruction 
are generally oriented perpendicular to the $(3\,\times\,1)$-ordered CoO$_2$ chains, possibly due to 
the incommensurability of $(2\,\times\,1)$ oxygen-reconstructed and $(3\,\times\,1)$-ordered TMO chains.  
The presence of oxygen-reconstructed areas without Co possibly indicates 
that the amount of deposited Co was slightly below one third of a ML. 
With a density about 0.03 nm$^{-2}$ the most frequent defects are point-like protrusions on the TMO chains. 
Their height amounts to about 80\,pm, consistent with typical values for transition metal adatoms. 
Therefore, we assume that these protrusions are caused by either excess Co 
or Ir atoms which were expelled from the surface layer but remained on the CoO$_2$ chains.
Furthermore, a few depressions in the chains can be recognized (one of which is marked by the green arrow).  
Since these depressions are centered where one would expect a maximum in a periodic chain in the absence of a defect, 
it appears reasonable to preliminarily assign them to Co vacancies.  
Both types of defects with similar characteristics will also appear for the other transition metals studied.  
In addition, we occasionally observe weak circular depressions (see blue arrow) 
which have also been observed on the bare Ir(001) and are assigned to sub-surface defects. 

To verify the structural properties of the CoO$_2$ on Ir(001) we measured a line profile perpendicular to the chains 
in between the two black arrows in the bottom right corner of the STM image displayed in Fig.\,\ref{Fig:CoO-Ir}(b). 
It is plotted at the bottom of Fig.\,\ref{Fig:CoO-Ir}(b). 
The periodicity of $(828\pm30)$\,pm agrees well with the expected value of $3a_\textrm{Ir} = 813$\,pm.\cite{Arblaster2010} 
Additionally, the atomic resolution scan shown in Fig.\,\ref{Fig:CoO-Ir}(c) 
and the corresponding line section shown in the bottom panel 
also confirm the $\times\,1$ periodicity with an atomic distance of $(278\pm10)$\,pm along the TMO chains.

After structural analysis we prepared magnetic Cr/W tips and Fe/W tips for SP-STM measurements. 
As documented in the Supplementary Information, these out-of-plane and in-plane sensitive tips 
were thoroughly tested by imaging the domain and domain wall structure the Fe DL on W(110), respectively.\cite{Sup} 
Although these tests clearly confirmed the magnetic imaging capabilities of our SP-STM tips 
before and after the measurements on the CoO$_2$ chains on Ir(001),
we never observed any magnetic contrast on the $(3\,\times\,1)$ structure of CoO$_2$ chains (not shown).  
This result is not necessarily in contradiction with the ferromagnetic order predicted in Ref.\,\onlinecite{Ferstl2016} 
because we have to keep in mind that the imaging of magnetic spin structures by SP-STM relies on the existence of domains 
or other spatial variations of the projection of the local sample magnetization onto the fixed magnetization of the tip. 
Therefore, the ferromagnetic domains could remain undetected if their size was much larger than the scan size.  
In this context future spatially averaging techniques, such as the magneto-optical Kerr effect, might be useful to clarify this open issue.


\subsubsection{\texorpdfstring{FeO$_{2}$/Ir(001)}{FeO2/Ir(001)}}
\label{Sect:FeO2-Ir}
\begin{figure}[b]
	\includegraphics[width=0.65\columnwidth]{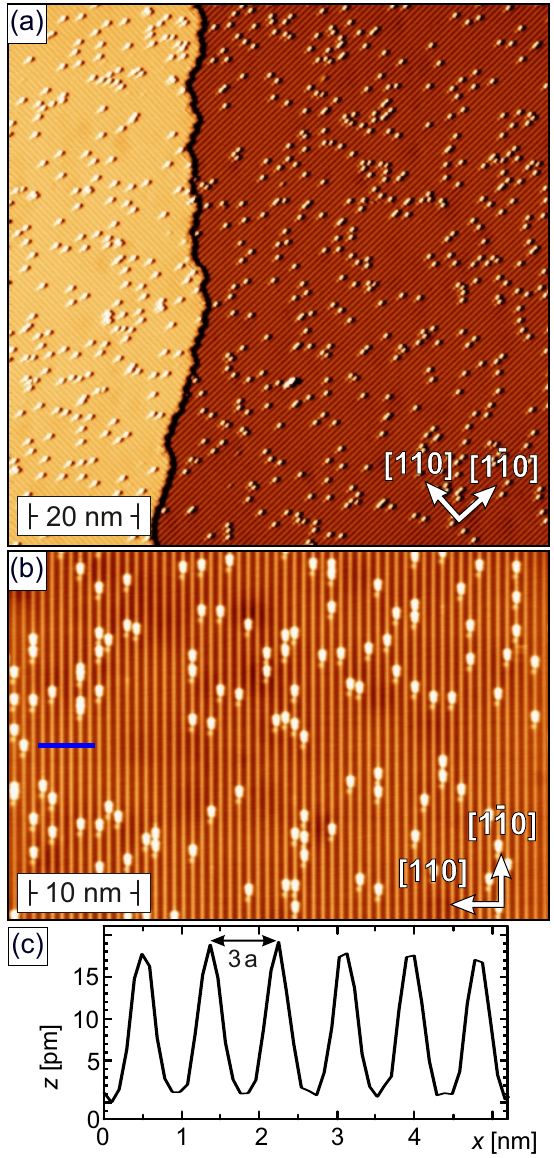}%
		\caption{(a) Overview scan of FeO$_2$/Ir(001) with chains running into $[110]$ direction on the upper 
				and $[1\overline{1}0]$ direction on the lower terrace. 
				(b) Higher resolution scan on the lower terrace. 
				(c) Line profile along the blue line in (b).  The periodicity of the stripes amounts to $(855\pm50)$\,pm.      
				Scan parameters: (a) \& (b) $U = -1$\,V, $I = 300$\,pA, non-magnetic W tip.}
		\label{Fig:FeO-Ir}
\end{figure}
The next transition metal element with one electron less in the $3d$ shell is Fe.  
It has been predicted that FeO$_2$ on Ir(001) exhibits an AFM order along the chains.\cite{Ferstl2016}
Just like for the preparation of CoO$_2$ chains described in Sect.\,\ref{Sect:CoO2-Ir}, 
Fe was evaporated at room temperature on the $(2\,\times\,1)$-reconstructed O/Ir(001) surface 
and immediately annealed at $T_{\rm ann} \approx 970$\,K 
in an oxygen atmosphere ($p_{{\rm O}_2} = 1 \times 10^{-7}$\,mbar). 
The overview scan in Fig.\,\ref{Fig:FeO-Ir}(a) shows two flat terraces separated by a monatomic substrate step edge. 
The absence of rectangular islands indicates that all Ir atoms 
expelled from the substrate had the chance to diffuse to step edges. 
Close inspection reveals that the chains are oriented along the $[110]$ direction on the upper (left) terrace 
whereas they are oriented in the $[1\overline{1}0]$ direction on the lower (right) terrace. 
The density of the point-like defects on top of the chains amount to about 0.06\,nm$^{-2}$.
Similar to what was discussed in the preceding section \ref{Sect:CoO2-Ir}, 
we believe that the most likely origin of these protrusions is a slight overdosing of the surface 
with Fe such that some atoms cannot be accommodated within the resulting Fe oxide chain structure.
We would like to note that the teardrop-shaped appearance of the protrusions 
is an imaging artifact caused by an unusual shape of the tip used in this experiment. 

\begin{figure}[tb]
	\includegraphics[width=0.9\columnwidth]{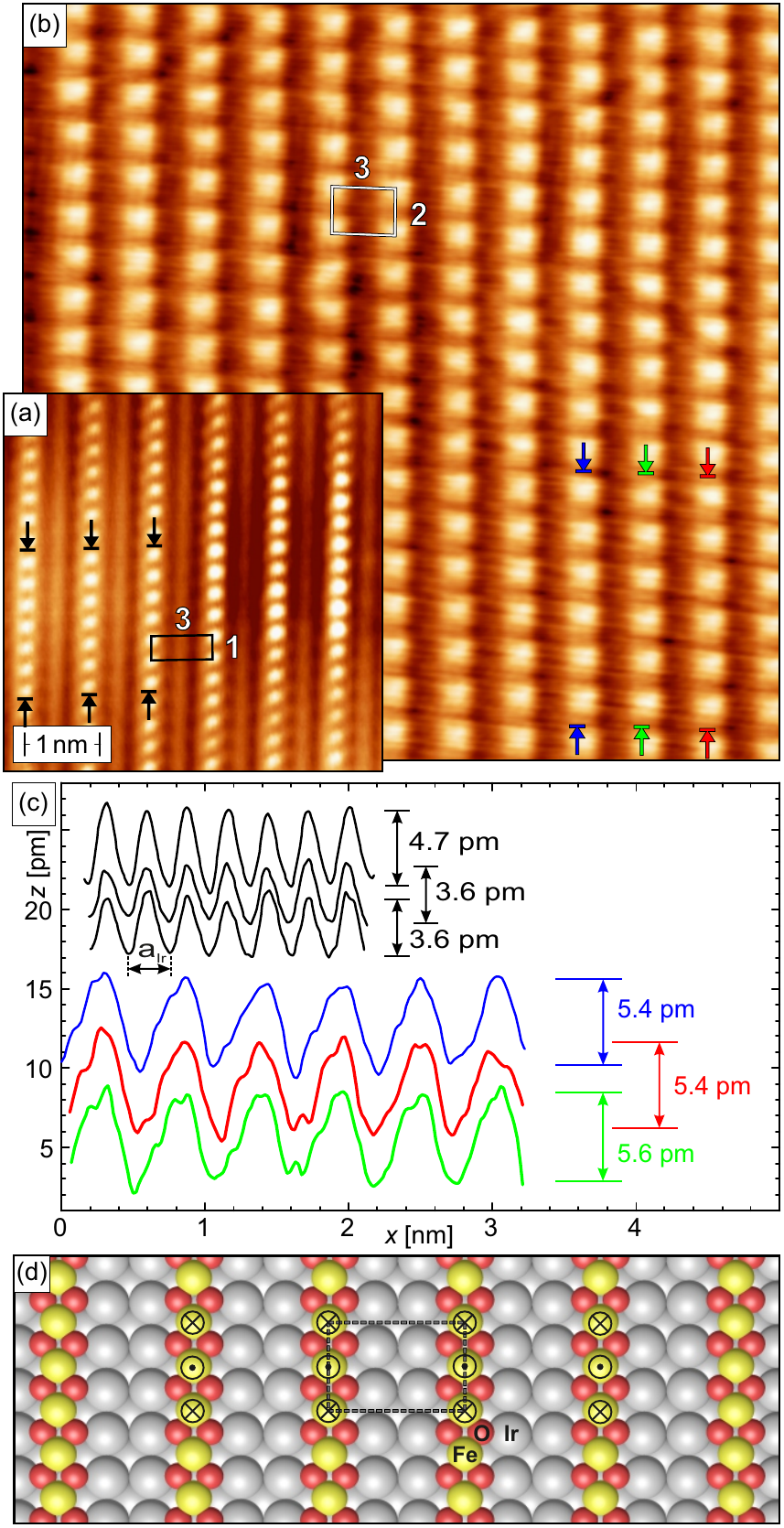}%
		\caption{(a) Atomic resolution scan of FeO$_2$ chains on Ir(001) obtained with a non-magnetic W tip. 
				The atoms along the chains are separated by one Ir lattice constant $a_\textrm{Ir} = 271$\,pm. 
				(b) SP-STM image obtained with an out-of-plane Cr/W tip. 
				Comparison to (a) reveals period along the chains lead corresponding to $2a_\textrm{Ir} = 542$\,pm
				resulting in a $(3\,\times\,2)$ magnetic unit cell. 
				(c) Line profiles measured along the chains at the position indicated by the arrows in (a) and (b). 
				(d) Schematic illustration of the $(3\,\times\,2$) collinear magnetic order. 
				While the magnetic coupling is AFM along the chains, the interaction of adjacent chains is FM.     
				Scan parameters: (a) $U = -10$\,mV, $I = 20$\,nA; (b) $U = 50$\,mV, $I = 5$\,nA. }
	\label{Fig:FeO-Ir-mag}
\end{figure}

For structural analysis of the FeO$_2$ chains a higher resolution scan on the lower terrace is shown in Fig.\,\ref{Fig:FeO-Ir}(b). 
The periodicity across the chains along the blue line is determined by the line profile shown in Fig.\,\ref{Fig:FeO-Ir}(c). 
Again we find periodicity of $(855\pm50)$\,pm which is in good agreement 
with the value expected for a $(3\,\times\,1)$ reconstruction, i.e., $3a_\textrm{Ir} = 813$\,pm. 
To complete the structural analysis an atomic resolution scan of FeO$_2$ is presented in Fig.\,\ref{Fig:FeO-Ir-mag}(a). 
Line sections drawn between the black arrows in Fig.\,\ref{Fig:FeO-Ir-mag}(a) which are presented in Fig.\,\ref{Fig:FeO-Ir-mag}(c)
(black lines) show a periodicity of $(278\pm10)$\,pm, consistent with the Ir lattice constant $a_\textrm{Ir} = 271$\,pm. 
Furthermore, an additional weak stripe in the center of the $(3\,\times\,1)$ unit cell 
in Fig.\,\ref{Fig:FeO-Ir-mag}(a) can be recognized in between the chains. 
A similar observation was reported by Ferstl {\em et al.}\cite{Ferstl2016} 
and assigned to an electronic signal of the Ir double-rows separating adjacent TMO chains.

The magnetic structure of this surface was investigated by SP-STM with an out-of-plane sensitive Cr-coated W tip.
The result is presented in Fig.\,\ref{Fig:FeO-Ir-mag}(b). 
Direct comparison with Fig.\,\ref{Fig:FeO-Ir-mag}(a) reveals a doubled period along the chains. 
This impression is corroborated by the line profiles presented in Fig.\,\ref{Fig:FeO-Ir-mag}(c) 
which have been taken in between the three pairs of colored arrows 
and clearly show a period $2a_\textrm{Ir} = 542$\,pm along the chains. 
The doubling is caused by the magnetoresistance effect which leads to a tunneling current 
which scales with the projection of the sample magnetization onto the tip magnetization (cf.\ Eq.\,\ref{Eq:SP-STM}).  
The most obvious explanation for the observed doubling of the period along the chain is an AFM coupling.
Furthermore, the maxima and minima of adjacent FeO$_2$ chains in Fig.\,\ref{Fig:FeO-Ir-mag}(b,c) 
are aligned, thereby clearly indicating a FM coupling across the chains. 
The fact that all magnetic line profiles show the same corrugation essentially excludes any non-collinear magnetic order, 
such as spin spirals or frustrated N{\'{e}}el spin states, but rather supports that FeO$_2$/Ir(001) exhibits collinear magnetism. 

Our experimental results obtained on FeO$_2$ chains on Ir(001) indicate a $(3\,\times\,2)$ magnetic unit cell 
which is marked as a box in Fig.\,\ref{Fig:FeO-Ir-mag}(b) and schematically illustrated in Fig.\,\ref{Fig:FeO-Ir-mag}(d). 
The AFM coupling observed along the chains is in agreement with recent theoretical predictions.\cite{Ferstl2016} 
To our opinion it is quite reasonable to assume that this AFM along the chains 
is caused by superexchange mediated by the fully occupied oxygen $2p$ orbitals.\cite{Anderson_SuperExchange}  
Although a much weaker AFM coupling across the chains was predicted in Ref.\,\onlinecite{Ferstl2016}, 
we experimentally observe spontaneous, permanent, and long-range FM inter-chain coupling at $T = 5$\,K.
At the moment we can only speculate why theory and experiment deviate.
Since adjacent TMO chains are separated by two non-magnetic Ir rows this order cannot be mediated by direct exchange. 
As we will point out below for MnO$_2$ chains on Ir(001), it appears that the strong spin-orbit interaction 
of the Ir substrate plays a decisive role for the formation of the magnetic ground state.  
Potentially, similar effects are also relevant for FeO$_2$/Ir(001).  
More advanced theoretical considerations will be necessary to fully elucidate the coupling mechanism. 

\subsubsection{\texorpdfstring{MnO$_{2}$/Ir(001)}{MnO2/Ir(001)}}
\label{Sect:MnO2-Ir}

\begin{figure}[tb]
	\includegraphics[width=0.65\columnwidth]{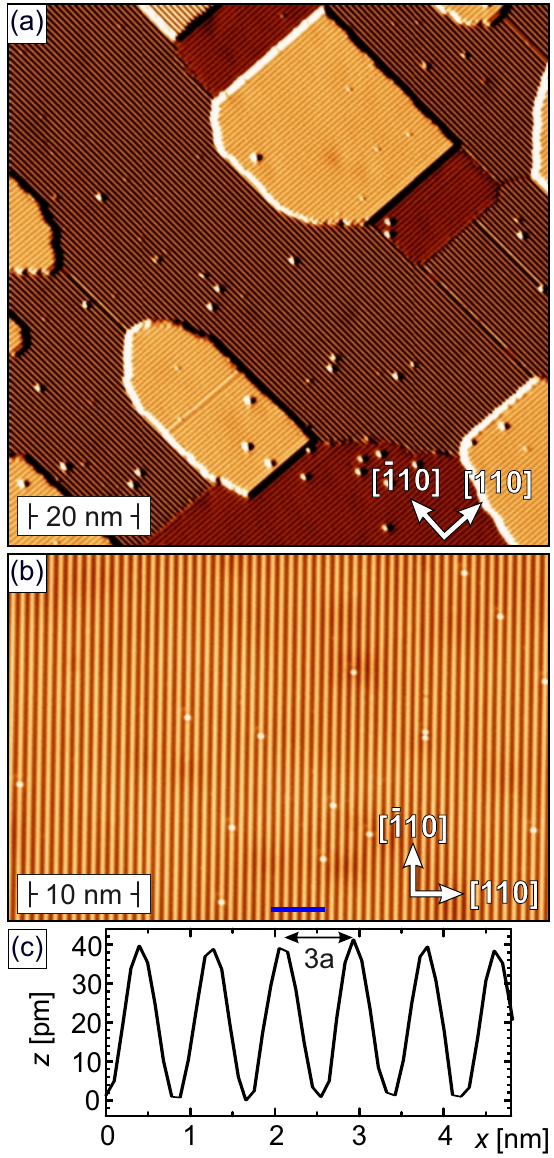}%
		\caption{(a) Overview scan of MnO$_2$ chains on Ir(001). Domains with stripes of the $(3\,\times\,1)$ structure 
				orientated along the $[\overline{1}10]$ and the $[110]$ direction can be recognized. 
				(b) Spin-averaged STM image of an atomically flat surface area showing a few defects only.  
				(c) Line profile measured along the blue line in (b) confirming an inter-chain distance of $3a_\textrm{Ir}$.     
				Scan parameters: $U = 1$\,V, $I = 300$\,pA, non-magnetic W tip.}
	\label{Fig:MnO-Ir}
\end{figure}

Our SP-STM results presented in the preceding section were in good agreement with DFT predictions 
regarding the intra-chain magnetic coupling,\cite{Ferstl2016} but also indicate 
that some aspect of the inter-chain coupling may not have been captured with sufficient accuracy. 
Since DFT qualitatively predicts the same intra-chain magnetic coupling for MnO$_2$/Ir(001) as for FeO$_2$ 
it is highly interesting to experimentally investigate also this sample system. 
The preparation of MnO$_2$ chains is very similar to the TMOs discussed so far. 
Also in this case Mn is deposited onto the oxidized Ir(001) surface at room temperature. 
To our experience the best surface quality can be achieved when choosing 
a slightly higher annealing temperature $T_{\rm ann} \approx 1020$\,K. 
This results in a sample topography with roughly rectangular islands, 
as exemplarily presented in the overview scan of Fig.\,\ref{Fig:MnO-Ir}(a). 
The $(3\,\times\,1)$-ordered MnO$_2$ chain structure covers almost the entire sample surface 
such that no regions with the oxygen-induced $(2\,\times\,1)$ reconstruction of Ir(001) remain visible.  
Typical $(3\,\times\,1)$ domains are about 20 to 40\,nm wide and often hundreds of nm long.  
Adjacent domains are separated by either orientational and anti-phase domain boundaries.

Similar to other TMO chains on Ir(001) we observe several defects which are located on top of the chains. 
For example, in Fig.\,\ref{Fig:MnO-Ir}(b) 12 single protrusions, one dimer, a vacancy in a MnO$_2$ row can be recognized.  
This corresponds to a defect density of only 0.015 nm$^{-2}$, well below what has been determined 
for CoO$_2$ and FeO$_2$ in Sect.\,\ref{Sect:CoO2-Ir} and \ref{Sect:FeO2-Ir}, respectively.  
The periodicity of the stripes can be determined by the line profile presented in Fig.\,\ref{Fig:MnO-Ir}(c). 
In agreement with the expected $(3\,\times\,1)$ structure it amounts to $(840 \pm 50)$\,pm.  

\begin{figure}[tb]
	\includegraphics[width=0.9\columnwidth]{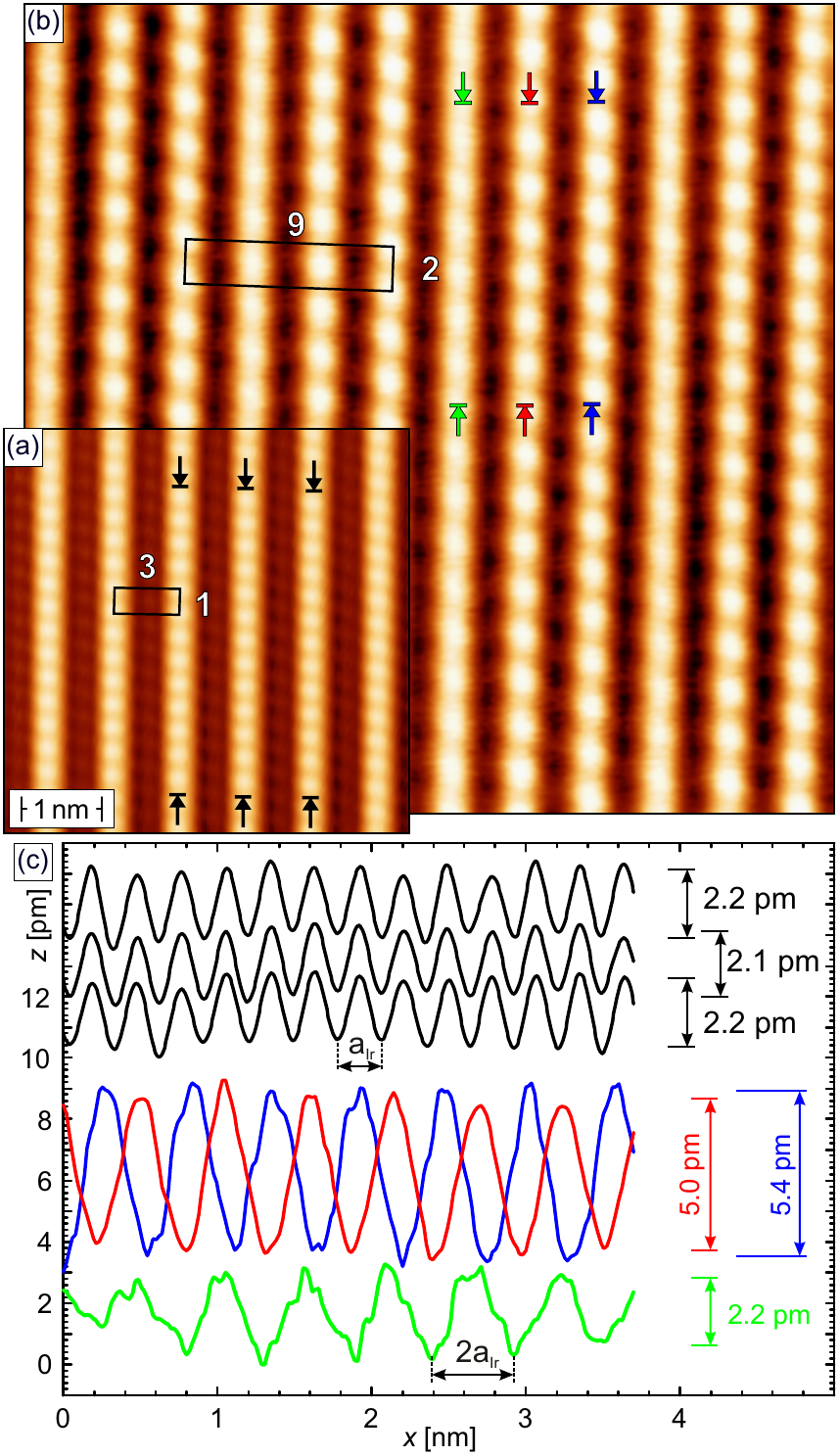}%
		\caption{(a) Atomic resolution scan of the MnO$_2$ chains on Ir(001) 
			measured with a non-magnetic W tip showing the $(3\,\times\,1)$ unit cell. 
			(b) SP-STM image of MnO$_2$/Ir(001) obtained with an in-plane sensitive Cr-coated W tip. 
			The period along the stripes corresponds to $2a_\textrm{Ir}$ indicating AFM order along the chain.
			The magnetic contrast of adjacent chains is modulated by phase and corrugation changes 
			resulting in a $(9\,\times\,2)$ magnetic unit cell. 
			(c) Line profile measured along three adjacent chains in (a) 
			without (black) and in (b) with magnetically sensitive tips (colored). 
			Scan parameters: (a) $U = -300$\,mV, $I = 3$\,nA; (b) $U = 200$\,mV, $I = 300$\,pA.}
	\label{Fig:MnO-Ir-mag}
\end{figure}
As shown in Fig.\,\ref{Fig:MnO-Ir-mag}(a) this $(3\,\times\,1)$ superstructure has been imaged with atomic resolution 
in a spin-averaging STM experiments by using a non-magnetic W tip.  
The corrugation and periodicity along the chains in Fig.\,\ref{Fig:MnO-Ir-mag}(a) 
can be determined from the line profiles presented in Fig.\,\ref{Fig:MnO-Ir-mag}(c) (black lines). 
These data sets clearly show that any chains exhibits a corrugation 
of $z \approx 2.2$\,pm only and a periodicity of $(287 \pm 20)$\,pm. 
In addition, Fig.\,\ref{Fig:MnO-Ir-mag}(a) also reveals a small corrugation 
in between the chains which can possibly be assigned to Ir pairs.

In Fig.\,\ref{Fig:MnO-Ir-mag}(b) we present spin-polarized measurements of MnO$_2$ chains on Ir(001) 
which were performed with an in-plane sensitive magnetic Cr/W tip. 
We would like to note that these results, which were obtained in a different experimental run 
on a newly prepared sample with a different SP-STM tip, qualitatively reproduce earlier results.\cite{NC}
Quantitative differences, such as a different corrugation, are attributed to the fact 
that the two tips probably differ in their spin polarization and quantization axes. 
Similar to the results on FeO$_2$ discussed in Sect.\,\ref{Sect:FeO2-Ir} the use of a magnetic probe tip 
leads to a doubling of the measured corrugation period which now corresponds to $2a_\textrm{Ir}$ along the chains. 
Furthermore, comparison of adjacent chains reveals a periodic variation of the corrugation. 
This can qualitatively be realized by comparing the chain marked with red arrows in Fig.\,\ref{Fig:MnO-Ir-mag}(b) 
with the chains on the left and right which are marked green and blue, respectively. 
The green chain on the left obviously exhibits a much weaker corrugation. 
In contrast, no obvious difference between the contrast strength of the red and blue line can be recognized, 
but the positions of maxima and minima of the chain marked with blue arrows are interchanged with respect to the red line. 

\begin{figure}[tb]
	\includegraphics[width=0.9\columnwidth]{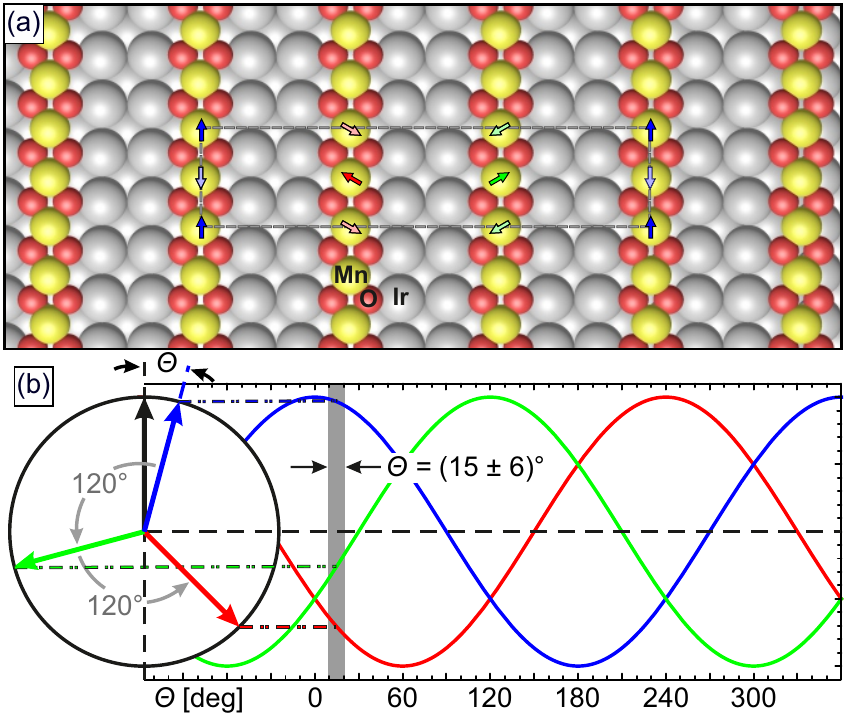}%
		\caption{(a) Schematic illustration of the $(9\,\times\,2)$ magnetic unit cell of MnO$_2$ chains on Ir(001). 
				The arrows indicate the spin direction of the corresponding Mn atoms. 
				(b) Sketch of the SP-STM signal expected for the proposed chiral $120^\circ$ spin spiral. 
				The corrugation and the phase shift observed in Fig.\,\ref{Fig:MnO-Ir-mag}(b) can be explained 
				by an angle $\Theta = (15 \pm 6)^\circ$ between tip polarization (black arrow) 
				and the TMO chain with highest corrugation (blue line).}
	\label{Fig:MnO-Ir-mod}
\end{figure}
The colored line profiles presented in Fig.\,\ref{Fig:MnO-Ir-mag}(c) allow for a more quantitative comparison.  
In agreement to our qualitative assessment the green chain exhibit the smallest corrugation of $2.2$\,pm  
whereas both, the blue and red line have a much higher corrugation of about 5\,pm. 
In addition, we can also recognize the phase shift of $a_\textrm{Ir}$  between the red and blue line. 
Detailed comparison shows that the blue chain has a slightly higher corrugation than the red chain. 
Together with the above-mentioned periodicity $2a_\textrm{Ir}$ along the chains 
these SP-STM observations suggest a $(9\,\times\,2)$ magnetic unit cell.

The experimental observations on MnO$_2$ chains can be explained by a spin structure 
as schematically illustrated in Fig.\,\ref{Fig:MnO-Ir-mod}(a). 
Each MnO$_2$ chain is AFM ordered whereby the easy axis always lies within the surface plane.  
The azimuthal orientation rotates by $120^\circ$ between adjacent MnO$_2$ chains 
such that the magnetic structure repeats every third chain.  
These two ingredients, i.e., the AFM order along and the tripled magnetic period perpendicular to the MnO$_2$ chains, 
result in a $(9\,\times\,2)$ magnetic unit cell as indicated in Fig.\,\ref{Fig:MnO-Ir-mag}(b).
As sketched in Fig.\,\ref{Fig:MnO-Ir-mod}(b), this magnetic structure can explain 
the corrugation and the phase relation observed by SP-STM in Fig.\,\ref{Fig:MnO-Ir-mag}(b,c). 
In this representation the black arrow represents the tip magnetization 
and the colored arrows represent the quantization axes of the respective chains. 

Proceeding from right to left in the $(9\,\times\,2)$ magnetic unit cell in Fig.\,\ref{Fig:MnO-Ir-mod}(a) 
we can recognize that the magnetic structure is governed by an in-plane counterclockwise spin spiral 
with a rotation angle of $120^\circ$ between adjacent chains.  
Since the blue arrow exhibits the largest projection onto the black arrow it will---according to 
Eq.\,\ref{Eq:SP-STM}---give the strongest magnetic contrast in SP-STM measurements, 
symbolized by the dot-dashed line in Fig.\,\ref{Fig:MnO-Ir-mod}(a).  
As a result of the spin spiral the green and red arrows are aligned antiparallel with respect to the tip. 
In case of an AFM intra-chain coupling this will unavoidably lead to a phase shift 
between the blue spin chain as compared to the green and red spin chain. 
Based on this representation we can nicely reproduce the corrugations extracted from Fig.\,\ref{Fig:MnO-Ir-mag}(c) 
by an angle between the blue chain and the tip polarization of $\theta = 15 \pm 6^\circ$, 
indicated by a grey region in Fig.\,\ref{Fig:MnO-Ir-mod}(b). 

\begin{figure*}[t]
	\includegraphics[width=0.99\textwidth]{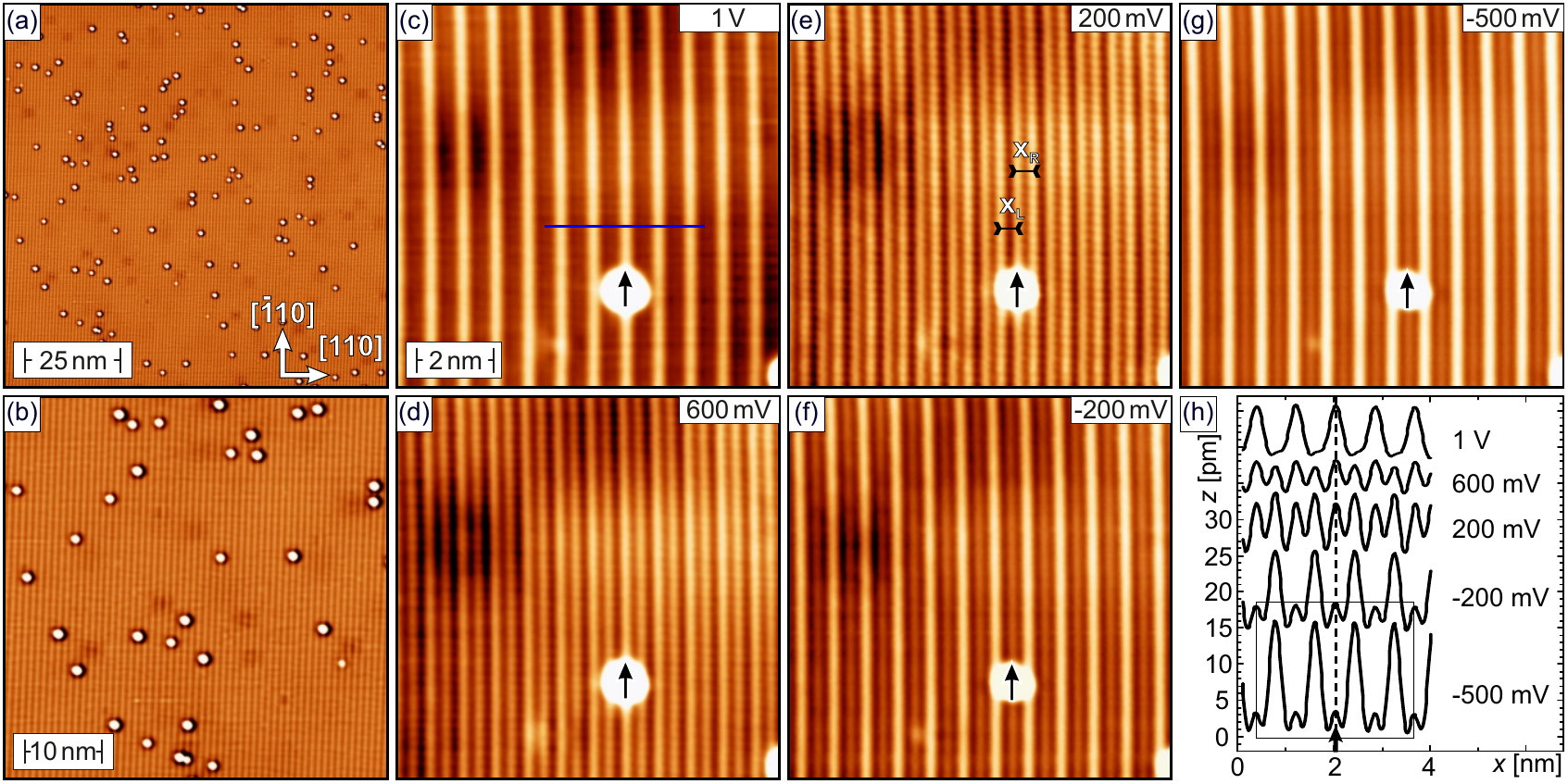}%
		\caption{(a) Overview and (b) higher resolution scan on CrO$_2$ chains on Ir(001). 
			(c)--(g) Bias-dependent STM images showing the same sample area.  
			A cluster (marked by an arrow) serves as a position marker.  
			A corrugation reversal is observed between panels (e) and (f), i.e., between $U = 200$\,mV and $U = -200$\,mV. 
			(h) Line profiles measured along lines corresponding to the blue line in (c).
			Scan parameters: (a),(b) $U = 1$\,V, $I = 300$\,pA; (d) $I = 2$\,nA; (d),(g) $I = 1.5$\,nA; (e),(f) $I = 0.5$\,nA.  }
	\label{Fig:CrO-Ir}
\end{figure*}

In a recently published paper the observation of a chiral $120^\circ$ spin spiral 
which is mediated across two atomic rows of the non-magnetic Ir(001) substrate 
was explained by a Dzyaloshinskii-Moriya type (DM) enhancement of the RKKY interaction.\cite{NC} 
This interaction, which has already been predicted in 1980\cite{FL1980}, 
leads to a magnetic coupling which is at the same time indirect and chiral. 
Evidence for DM-enhanced indirect magnetic coupling had previously been observed on some surface-deposited 
clusters\cite{2015Dupe,Khajetoorians2016,Bouaziz2017,Hermenau2017} and Dy/Y superlattices.\cite{GCY2008}  
Indeed, state-of-the-art density functional theory calculations could successfully reproduce not only the antiferromagnetic coupling 
along the chains but also explain the period of the chiral magnetic structure across the chains.\cite{NC} 
However, the experimental data suggest a spin spiral which predominantly rotates within the surface plane, 
corresponding to a Dzyaloshinskii-Moriya vector $\mathbf{D}$ oriented along the surface normal. 
This surprising result cannot be explained within the structural model deduced 
from quantitative low-energy electron diffraction data,\cite{Ferstl2016} 
which would result in a $\mathbf{D}$ vector within the surface.  
The discrepancy might be caused by yet unconsidered effects of structural domain boundaries or relaxation effects. 
Additional DFT calculations and field-dependent SP-STM measurements will be necessary to fully resolve this open question.

\subsubsection{\texorpdfstring{CrO$_{2}$/Ir(001)}{CrO2/Ir(001)}}
\label{Sect:CrO2-Ir}
Motivated by the observation of an unexpected non-collinear magnetism 
in $(3\,\times\,1)$-ordered MnO$_2$ chains on Ir(001) we tried to extent the series 
of self-organized TMO chains to Cr which is the left neighbor of Mn in the periodic table of elements. 
Since the growth of CrO$_2$ chains was not part of the study 
by Ferstl {\em et al.}\cite{Ferstl2016} it will be discussed here in detail.
The preparation follows the principle of the known TMO chains and comprises 
the evaporation of Cr onto the $(2\,\times\,1)$ O/Ir(001) surface at room temperature. 
Subsequently, the sample was annealed in an oxygen atmosphere 
($p_{\rm{O_{2}}} = 1 \times 10^{-8}$\,mbar) at a temperature $T_{\rm ann} \approx 1070$\,K. 

The resulting sample topography and its structural analysis are shown in Fig.\,\ref{Fig:CrO-Ir}. 
Already in the overview scan presented in Fig.\,\ref{Fig:CrO-Ir}(a) 
we can recognize stripes orientated along the $[\overline{1}10]$ direction. 
We count 145 point-like defects with a typical height of several 10 up to 100\,pm on top of the stripes, 
corresponding to a defect density below 0.015 per nm$^2$.
Moreover, some protrusions are visible that we already identified on clean $(5\,\times\,1)$ reconstructed Ir(001) surfaces. 
The periodic arrangement of the stripes with an inter-stripe distance 
of $(846\pm40)$\,pm is shown at higher resolution in Fig.\,\ref{Fig:CrO-Ir}(b).
This value is comparable to the inter-chain distance observed on the other quasi one-dimensional TMO systems 
and consistent with the $(3\,\times\,1)$ structural unit cell. 

To study the electronic properties of this stripe pattern 
we performed the bias-dependent measurements presented in Fig.\,\ref{Fig:CrO-Ir}(c)--(f). 
All scans show the same sample area as can be recognized by the cluster of unknown origin. 
An arrow marks the central region of this cluster and serves as a reference point during the following comparison. 
At a bias voltage $U = 1$\,V, Fig.\,\ref{Fig:CrO-Ir}(c), the arrow is aligned with the top of a stripe (A-stripe). 
The next stripe to the right can be found about 1\,nm apart just right of the bright cluster.
When the bias voltage is reduced to $U = 600$\,mV, Fig.\,\ref{Fig:CrO-Ir}(d), 
another stripe (B-stripe) appears between the A-stripes.  
Interestingly, the corrugation is not mirror symmetric with respect to the plane 
defined by the $[{\bar 1}10]$ direction, i.e., along the stripes, and the (001) surface normal.  
This can best be recognized in the line sections shown in Fig.\,\ref{Fig:CrO-Ir}(h).  
The data have been extracted from the images shown in Fig.\,\ref{Fig:CrO-Ir}(c)--(f) 
along traces corresponding to the blue line in panel (c). 
Again the arrow (and the hatched line) indicates the central position of the bright cluster.  
At $U = 600$\,mV the two minima left and right of the hatched line clearly exhibit different depth, 
with the left minimum being about 1\,pm deeper than the right minimum.  
This trend becomes even stronger as the bias voltage is changed to $U = 200$\,meV.  
Furthermore, when decreasing the bias voltage further to $U = -200$\,meV, Fig.\,\ref{Fig:CrO-Ir}(f), 
and $U = -500$\,meV, Fig.\,\ref{Fig:CrO-Ir}(g), we observe 
that the corrugation of the B-stripes becomes much larger than that of the A-stripes. 
For example, at $U = -500$\,meV the corrugation of B-stripes reaches $(15\pm1)$\,pm, 
whereas the corrugation of the A-stripes only amounts to $(2.6\pm0.3)$\,pm.

Our bias-dependent experimental results on CrO$_2$/Ir(001) are strikingly different 
as compared to similar results on the other TMO chains. 
We find an asymmetric corrugation and A- and B-stripes which are the dominant features 
in topographic STM measurements performed at $U > 200$\,meV and $U \leq -200$\,meV, respectively. 
Furthermore, the change in corrugation is more pronounced for B-stripes. 
Combined with the fact that the B-stripes almost vanish for $U = 1$\,V 
we suppose that A-stripes correspond to the CrO$_2$ chains
whereas the B-stripes are assigned to the intermediate Ir double row. 
As detailed in Ref.\,\onlinecite{Sup} we performed SP-STM experiments on CrO$_2$/Ir(001) 
with both, out-of-plane sensitive Cr/W and in-plane sensitive Fe/W tips. 
These measurements confirmed the structural $(3\,\times\,1)$ unit cell of the CrO$_2$ chains 
but did not provide any hint of a magnetically ordered state. 
       
\subsection{TMO chains on Pt(001)}
\label{sect:ResTMO_Pt}
\subsubsection{CoO$_{2}$/Pt(001)}
\label{Sect:CoO2-Pt}
Our SP-STM experiments of TMO chains on Ir(001) presented so far revealed a wide variety of spin structures. 
Whereas we could not detect any magnetic signal on CoO$_2$ chains, suggesting a non-magnetic state, 
for other $3d$ elements we found collinear (anti)ferromagnetic (Fe) and a helical spin spiral order (Mn). 
It appears to be an obvious idea to extend the search for similar indirect DM-enhanced RKKY interactions 
by growing TMO chains on other heavy noble metals which also crystallize in the fcc structure. 
One candidate material is Pt, the direct neighbor of Ir in the periodic table of elements.  
In fact, it has already been shown that CoO$_2$ chains can be grown by similar procedures on Pt(001).\cite{Ferstl2017}

\begin{figure}[t]
	\includegraphics[width=\columnwidth]{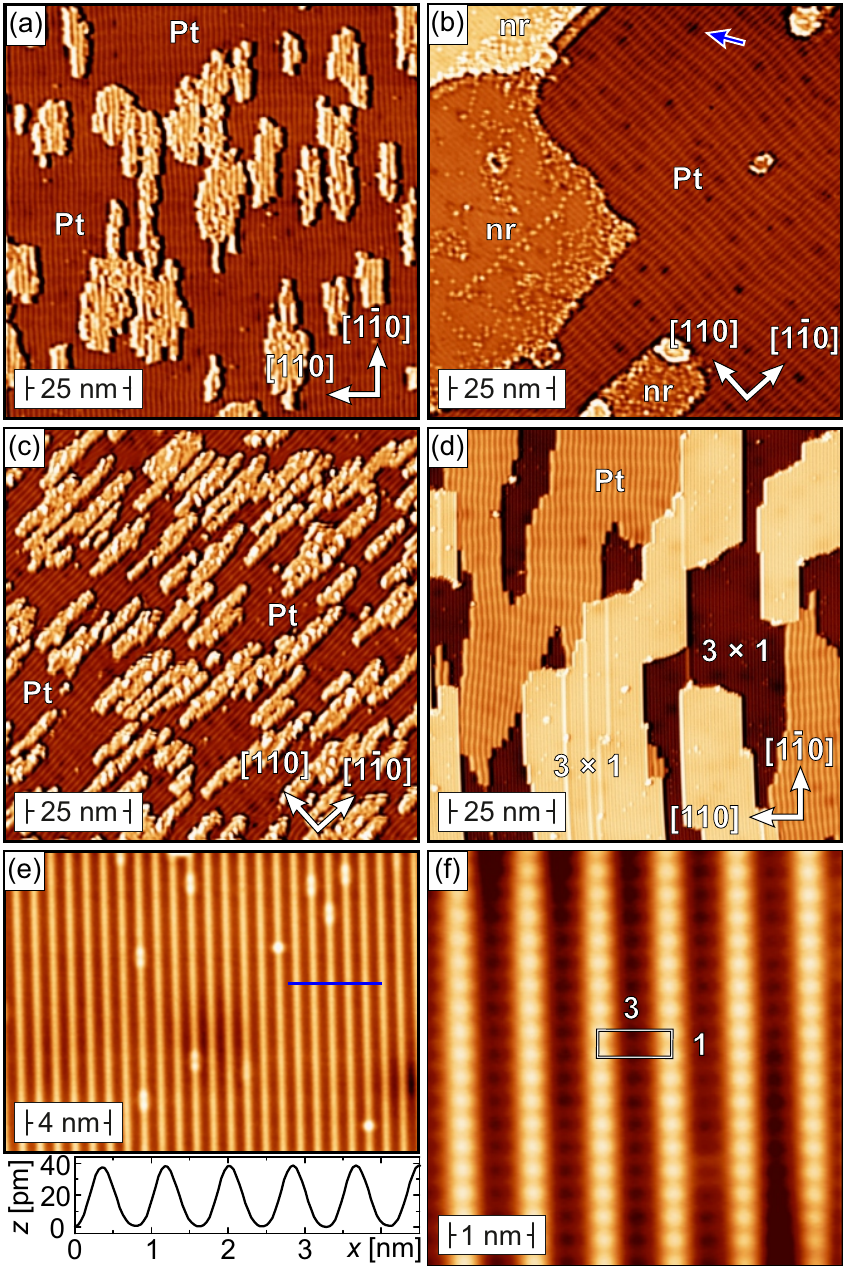}%
	\caption{Growth of CoO$_2$ chains on Pt(001). 
		(a) Topography of Co/Pt(001) after evaporation on the warm substrate. 
		(b) After oxidation of sample (a) no $(3\,\times\,1)$ structure is observed. 
		(c) Topography of Co/Pt(001) after evaporation on the cold substrate. 
		(d) After oxidation of sample (c) a coexistence of Pt reconstruction and domains of CoO$_2$ chains is observed. 
		(e) Higher resolution scan of the chains with line profile across the stripes. 
		(f) Atomic resolution scan with a non-magnetic W tip confirms the structural $(3\,\times\,1)$ unit cell.  
		Scan parameters: (a)--(d) $U = 1$\,V, $I = 300$\,pA; 
					(e) $U = -100$\,mV, $I = 1.5$\,nA; (e) $U = 50$\,mV, $I = 5$\,nA, 
					all data measured with a non-magnetic W tip.}
		\label{Fig:CoO-Pt}
\end{figure}
Since the Pt(001) reconstruction is stable against oxidation we first studied the growth of Co on the clean surface.\cite{Ferstl2017} 
In our STM images the complex hexagonal reconstructions of clean Pt(001)  
shows up as stripes with a periodicity of $(1.36 \pm 15)$\,nm.\cite{PhysRevB.94.195406} 
Fig.\,\ref{Fig:CoO-Pt}(a) shows the topography of a sample prepared by Co evaporation onto the warm surface 
resulting in the formation of $(163 \pm 15)$\,pm high islands which are elongated along the $[1\overline{1}0]$ direction. 
This orientation correlates with the stripe direction of the Pt(001) surface 
and indicates preferential diffusion along the trenches of the reconstruction. 

Annealing this sample in an oxygen atmosphere ($p_{{\rm O}_2} = 1 \times 10^{-7}$\,mbar) 
at temperature $T_{\rm ann} \approx 920$\,K results in the topography shown in Fig.\,\ref{Fig:CoO-Pt}(b). 
Two surface areas can be distinguished in these data which were measured with non-magnetic W tips.   
In the upper right of Fig.\,\ref{Fig:CoO-Pt}(b) we again recognize 
the clean Pt(001) surface which---similar to our observations on clean Ir(001) 
[cf.\ blue arrow in Fig.\,\ref{Fig:CoO-Pt}(b)]---exhibits occasional circular depressions (density $\approx 0.02$\,nm$^{-2}$). 
The persistence of this reconstruction confirms the stability of the Pt(001) surface against oxidation.  
The remaining part of the surface is covered by islands without any oxygen-induced reconstruction (nr). 
Together with the unusual step height of $(90 \pm 15)$\,pm 
we interpret this as evidence for the formation of a CoPt surface alloy.

In a second attempt we initially cooled the clean Pt(001) surface in the LT-STM down to $T \approx 5.5$\,K. 
Subsequently the crystal was transferred to the preparation chamber for Co deposition onto the cold surface. 
Comparison of the resulting surface topography, Fig.\,\ref{Fig:CoO-Pt}(c), 
with the Co film grown at elevated temperature, Fig.\,\ref{Fig:CoO-Pt}(a), 
shows that low-temperature growth leads to smaller Co islands and the nucleation of very few second layer Co islands. 
In contrast to the Co islands shown in Fig.\,\ref{Fig:CoO-Pt}(a) these low-temperature--deposited islands exhibit no stripes. 
Annealing this sample at the same parameters, $T_{\rm ann} \approx 920$\,K, 
leads to the topography presented in Fig.\,\ref{Fig:CoO-Pt}(d), 
which is comparable to the earlier results of Ref.\,\onlinecite{Ferstl2017}. 
We can distinguish areas of reconstructed clean Pt(001) from extended and well-ordered regions 
showing the $(3\,\times\,1)$ structure which is characteristic for CoO$_2$ chains. 
The step height between domains of the same structure amounts to $(199 \pm 10)$\,pm, 
in perfect agreement with the value expected for Pt(001). 
To complete the structural analysis we performed a higher resolution scan which are presented in Fig.\,\ref{Fig:CoO-Pt}(e). 
Again point-like defects are located on the stripes. 
The line profile taken at the position of the blue line confirms the periodicity of $(830 \pm 20)$\,pm. 
The inner structure of the stripes can be resolved by the atomic resolution image shown in Fig.\,\ref{Fig:CoO-Pt}(f).
Along the chains we measure a periodicity of $(278 \pm 15)$\,pm which agrees well 
with the cubic lattice vector expected for Pt, $a_\textrm{Pt} = 277$\,pm.\cite{Morgan1968}

After confirmation of the $(3\,\times\,1)$ structure of CoO$_2$ chains on Pt(001)\cite{Ferstl2017} 
we attempted to resolve the spin or magnetic domain structure 
by means of SP-STM experiments using out-of-plane and in-plane polarized tips. 
However, these magnetically sensitive experiments\cite{Sup} showed neither a hint 
of the theoretically proposed AFM structure\cite{Ferstl2017} 
nor could we detect any hint of other magnetically ordered states.

\subsubsection{FeO$_{2}$/Pt(001)}
\label{Sect:FeO2-Pt}
\begin{figure}[t]
	\includegraphics[width=0.99\columnwidth]{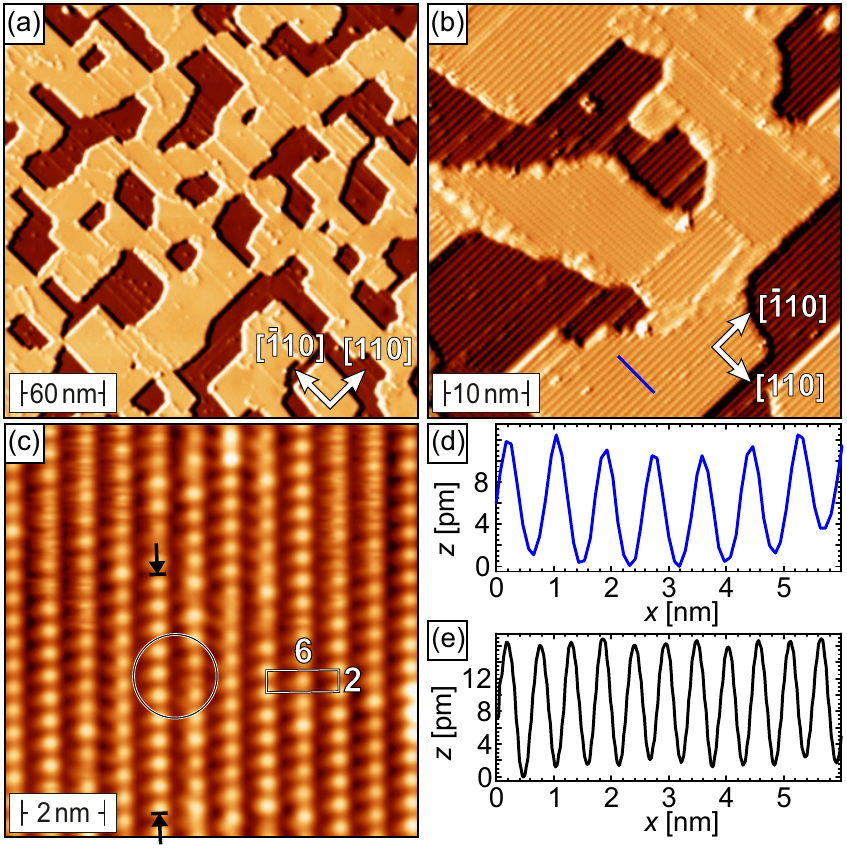}%
		\caption{(a) Overview scan of oxidized Fe on Pt(001). 
				(b) A higher resolution scan shows that terraces and islands exhibit a stripe pattern 
				with the expected $3a_\textrm{Pt}$ periodicity. 
				(c) Atomic scale image of oxidized Fe on Pt(001). 
				Along the stripes the corrugation has a period of twice the atomic lattice vector. 
				Within the circle the order across the stripes changes from aligned to a shifted position. 
				(d) Line profiles across (top) and along the stripes (bottom) at the positions of the blue line in (b) 
				and at the position of the arrows in (c).
				Scan parameters: (a),(b) $U = 1$\,V, $I = 300$\,pA; (c) $U = 100$\,mV, $I = 1$\,nA, non-magnetic W tip.   }
	\label{Fig:FeO-Pt}
\end{figure}

For the preparation of FeO$_2$ chains on Pt(001) we followed the same procedure as for Co, 
i.e., we started with the growth of Fe on the clean reconstructed Pt(001) surface. 
Indeed, Fe evaporation onto the warm Pt(001) substrate leads to a surface morphology very similar to Co/Pt(001) 
[cf.\ \ref{Fig:CoO-Pt}(a,b), results for Fe not shown here] which we interpret as evidence for alloying with the substrate.
To overcome this issue Fe was evaporated onto the cold sample and immediately oxidized at $T \approx 870$\,K. 
The resulting topography which is presented in the overview scan of Fig.\,\ref{Fig:FeO-Pt}(a) 
(scan size: $300$\,nm\,$\times\,300$\,nm) shows plateaus and valleys with a typical lateral size of about 50\,nm. 
They are separated by $(199 \pm 15)$\,pm high step edges 
preferentially oriented along high symmetry directions of the substrate. 
Scanning a similar surface area at higher resolution, Fig.\,\ref{Fig:FeO-Pt}(b), reveals a stripe pattern 
with a periodicity of $(844 \pm 40)$\,pm, as determined from the line profile in Fig.\,\ref{Fig:FeO-Pt}(d),
in agreement with the $(3\,\times\,1)$ reconstruction expected for FeO$_2$ chains on Pt(001). 
Regions with stripes oriented in the $[\overline{1}10]$ or the $[110]$ directions can be recognized.
   
Surprisingly, atomic scale images performed with non-magnetic tips within a single domain 
are inconsistent with the expected structural $(3\,\times\,1)$ unit cell. 
This can clearly be recognized by the image presented in Fig.\,\ref{Fig:FeO-Pt}(c) 
and the line section shown Fig.\,\ref{Fig:FeO-Pt}(e) which was measured in between the two black arrows.
The periodicity extracted from this line profile  along the chains amounts to $(543 \pm 25)$\,pm. 
This value is in good agreement with twice the lattice constant of Pt, $2a_\textrm{Pt}$.  
Furthermore, the maxima of adjacent stripes are out of phase in most cases.  
Therefore, our spin-averaged experimental data suggest the existence 
of a $(6\,\times\,2)$ unit cell for the oxidized Fe on Pt(001) [see white box in Fig.\,\ref{Fig:FeO-Pt}(c)].
Since the positions of maxima and minima in a given chain are not completely static 
but slowly fluctuate, occasional exceptions from this rule can be observed.  
For example, the chain in the right part of the white circle in Fig.\,\ref{Fig:FeO-Pt}(c) shows such a switching event. 
Whereas the two chains marked by the circle are in phase in the lower part of the image, 
a phase shifts by $a_\textrm{Pt}$ in the right chain causes that the two chains are out of phase in the upper part of the image. 
Since the measurement of Fig.\,\ref{Fig:FeO-Pt}(c) was executed with a non-magnetic tip
and since the use of out-of-plane or in plane sensitive magnetic probe tips didn't result in any additional contrast,
we have to conclude that the observed reconstruction has no magnetic but either a structural or an electronic origin. 
One possible explanation could be a Peierls instability due to a metal--isolator transition at low temperatures.\cite{Houselt2008} 
Further investigations are required to figure out the origin of the iron oxide phase.

\subsubsection{\texorpdfstring{MnO$_{2}$/Pt(001)}{MnO2/Pt(001)}}
\label{Sect:MnO2-Pt}

\begin{figure}[tb]
	\includegraphics[width=0.7\columnwidth]{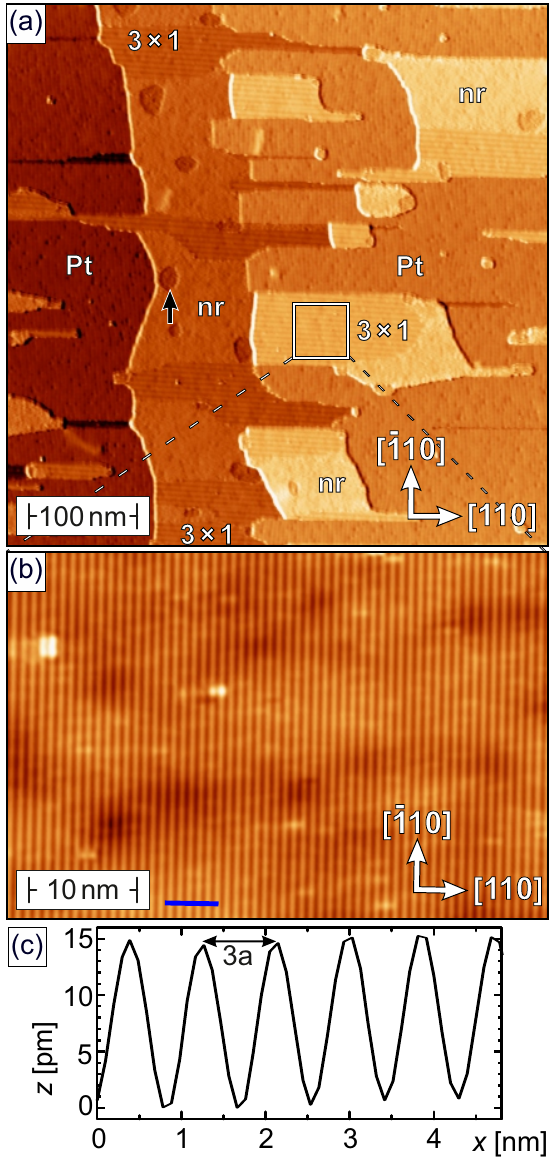}%
		\caption{(a) Overview scan of Pt(001) after Mn deposition and subsequent annealing in oxygen. 
			In addition to reconstructed clean Pt(001) one also observes disordered (nr) 
			and $(3\,\times\,1)$-reconstructed surface regions. 
			(b) Higher resolution scan at the position of the stripes. 
			(c) Line profile measured along the blue line in (b) confirming the $3a_\textrm{Pt}$ periodicity.    
			Scan parameters: $U = 1$\,V, $I = 300$\,pA, non-magnetic W tip.  }
	\label{Fig:MnO-Pt}
\end{figure}

Finally we investigated MnO$_2$ chains on Pt(001). 
In a first attempt we evaporated Mn while the Pt(001) substrate was held at room temperature 
and subsequently oxidized the sample at an oxygen pressure $p_{{\rm O}_2} = 1\times10^{-7}$\,mbar at $T \approx 920$\,K. 
Unfortunately, this preparation procedure resulted in very small TMO domains with size of $(10...20)$\,nm only.
In order to improve surface homogeneity we reduced the oxygen pressure to $p_{{\rm O}_2} = 1\times10^{-8}$\,mbar. 
The resulting topography measured with a non-magnetic W tip is shown in Fig.\,\ref{Fig:MnO-Pt}(a). 
A significant fraction of the surface still exhibits the familiar reconstruction of the clean Pt(001) surface. 
In addition, two other kinds of surfaces could be identified. 
The first one no longer shows any reconstruction (not reconstructed; nr) but is relatively rough. 
Similar findings have been reported for Cu/Ir(001) and were assigned to alloying.\cite{Gilarowski2000} 
Second, we found $(3\,\times\,1)$-reconstructed areas of MnO$_2$. 
For structural determination a higher resolution scan measured 
on a homogeneously striped domain is shown in Fig.\,\ref{Fig:MnO-Pt}(b). 
Adjacent stripes are separated by $(866 \pm 45)$\,pm, 
as determined from the corresponding line profile in Fig.\,\ref{Fig:MnO-Pt}(c).
We would like to emphasize that this particular structural domain exhibits a lateral size well above 50\,nm, 
thereby allowing for the clear identification even of large magnetic unit cells. 
Moreover the surface exhibits very few defects.

\begin{figure}[tb]
	\includegraphics[width=0.9\columnwidth]{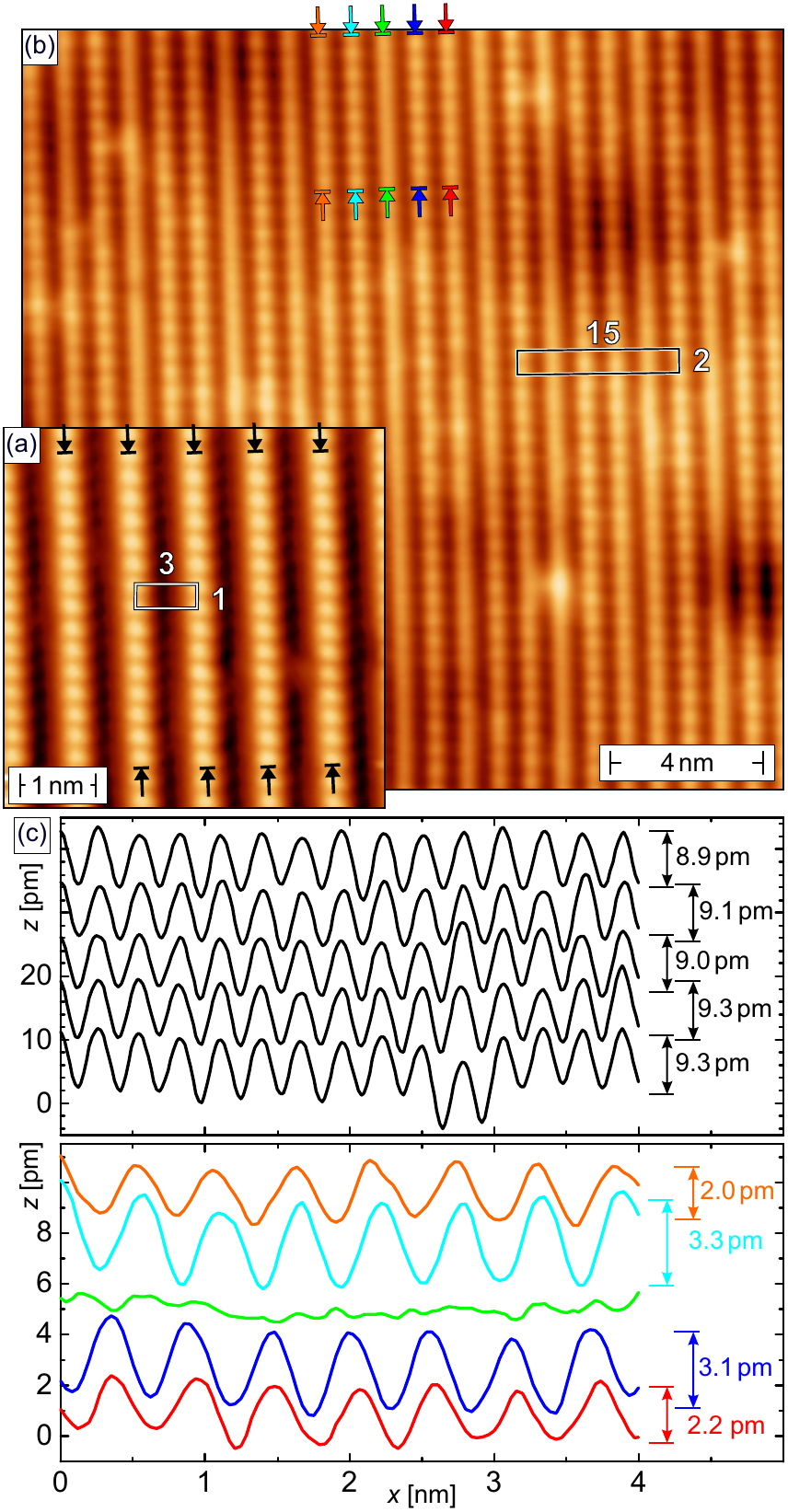}%
		\caption{(a) Atomic resolution scan of MnO$_2$ chains on Pt(001) performed with a W tip. 
			(b) Spin-polarized measurement taken with a Cr-coated W tip. 
			A periodic modulation of the chain-specific corrugation can clearly be seen. 
			(c) Line profiles measured along five adjacent chains marked by correspondingly colored arrows 
			as measured with a non-magnetic (top) and a magnetic tip (bottom). 
			Whereas a doubling of the period characteristic for an intra-chain AFM coupling is observed on most of the chains, 
			a pronounced modulation of the contrast is also visible across the chains. 
	Scan parameters: (a) $U = -5$\,mV, $I = 20$\,nA; (c) $U = 1$\,V, $I = 300$\,pA.  }
\label{Fig:MnO-Pt-mag}
\end{figure}
To exclude a behaviour similar to what we have observed on oxidized Fe on Pt(001) 
we performed a detailed analysis of the observed contrasts.  
For example, line profiles measured along the chains marked with black arrows in Fig.\,\ref{Fig:MnO-Pt-mag}(a) 
are presented in the upper part of Fig.\,\ref{Fig:MnO-Pt-mag}(c). 
With a corrugation between $8.9$ and $9.3$\,pm and an atomic spacing of $(281 \pm 19)$\,pm 
the data are in good agreement with a $(3\,\times\,1)$ structure of MnO$_2$ chains. 

Such a sample surface was investigated by means of SP-STM using out-of-plane sensitive Cr-coated W tips.  
The resulting SP-STM image presented in Fig.\,\ref{Fig:MnO-Pt-mag}(b) exhibits striking similarities 
if compared with the results obtained on MnO$_2$ chains on Ir(001) [cf.\ Fig.\,\ref{Fig:MnO-Ir-mag}(b)]. 
To analyze the spin-polarized measurement five line profiles which have been taken 
in between the colored arrows are plotted in the lower part of Fig.\,\ref{Fig:MnO-Pt-mag}(c). 
Comparison with the spin-averaged data (black) reveals that---identical to our results 
obtained on MnO$_2$/Ir(001)---a doubling of the period along the chains can be recognized. 
This is a clear sign for an AFM intra-chain coupling of the Mn atoms. 
In further analogy to the spin spiral system of MnO$_2$ on Ir(001) the magnetic corrugation 
measured on the respective chains for MnO$_2$/Pt(001) is not constant but modulates periodically. 
For example, the line profiles clearly show that the green chain exhibits the lowest, almost vanishing corrugation. 
The other four chains constitute two groups which can be distinguished on the basis of the phase of their magnetic contrast. 
Whenever the two upper colored line profiles (orange, light blue) exhibit a maximum, 
the lower line profiles (red, dark blue) show a minimum. 
Since all line profiles possess a periodicity $2a_\textrm{Pt}$ due to their AFM order along the chains, 
the phase difference shifts the magnetic contrast by $a_\textrm{Pt}$. 

\begin{figure}[tb]
	\includegraphics[width=0.9\columnwidth]{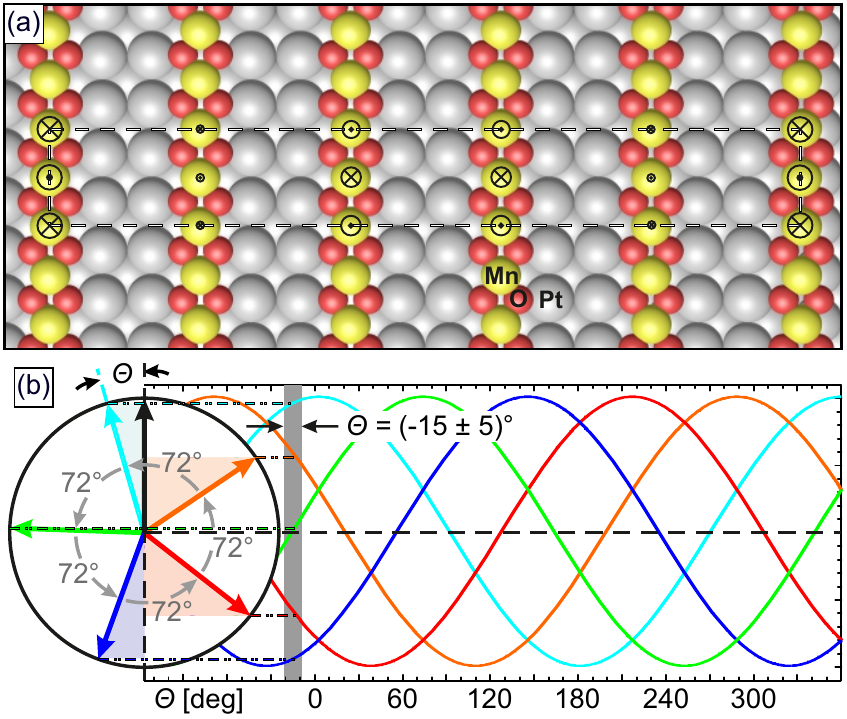}
		\caption{(a) Schematic representation of a $(15\,\times\,2)$ magnetic unit cell. 
		A cycloidal spin spiral with out-of-plane rotating spins. 
		The model in (b) is based on the magnetic corrugation in Fig.\,\ref{Fig:MnO-Pt-mag}(c). 
		From this a tilt $\Theta = (-15 \pm 5)^\circ$ of the tip polarization 
		with respect to the light blue chain with highest corrugation is seen.  }
\label{Fig:MnO-Pt-model}
\end{figure}
This behaviour can be explained by the model illustrated in Fig.\,\ref{Fig:MnO-Pt-model}(a). 
The coupling along the chains is assumed to be collinear AFM within each individual MnO$_2$ chain. 
We initially assume a periodic modulation by a commensurate $(15\,\times\,2)$ magnetic unit cell 
where the spin quantization axis of adjacent stripes rotates by $72^\circ$. 
Due to the dependence of the spin-polarized current on the angle between the tip magnetization 
and the spin quantization axis of the Mn atoms described by Eq.\,\ref{Eq:SP-STM} 
this leads to a characteristic behavior of the magnetic corrugation.  
It is highest for the TMO chains which exhibit the highest projection onto the tip magnetization vector 
symbolized by a black arrow in Fig.\,\ref{Fig:MnO-Pt-model}(b), i.e., the light and dark blue chains.  
Their opposite perpendicular orientations will result in a $\pi$ phase shift. 
A lower magnetic corrugation can be expected for the chains symbolized by orange and red arrows.  
The corrugation vanishes if the spin quantization  axis of a chain is perpendicular to the tip polarization (green arrow). 
As a consequence of this rotating spin structure two chains will exhibit a parallel (light blue, orange) 
and two an antiparallel (blue, red) alignment with respect to the tip, 
thereby explaining the phase shift of $a_\textrm{Pt}$ observed in Fig.\,\ref{Fig:MnO-Pt-mag}(b).  
On the basis of this model we are able to determine the angle between the magnetization direction of the tip 
and the light blue MnO$_2$ chain to $\Theta = (-15 \pm 5)^\circ$.  

SP-STM data measured with an in-plane sensitive Fe-coated W tip across a structural domain boundary 
indicate that the spin spiral possesses in-plane and out-of-plane contributions to the magnetization. \cite{Sup} 
A similar spin spiral driven by the Dzyaloshinskii-Moriya interaction has already been found 
by DFT calculations for MnO$_2$ chains on Ir(001).\cite{NC} 
In agreement with the so-called Moriya rules\cite{Mor1960} it is most natural 
to assume a cycloidal spin spiral with a Dzyaloshinskii-Moriya vector oriented along the chain direction.  
While the DMI is expected to be very similar for Ir and Pt due to their very similar atomic number, 
the higher occupation of the Pt $5d$ shell causes significant differences in the respective Fermi surfaces. 
Indeed, it has been found that the RKKY-mediated oscillation period for (001)-oriented magnetic superlattices 
is about 50\% longer for Co/Pt\cite{Mor2004} than for Fe/Ir,\cite{Stoeffler2004}   
in reasonable agreement with our observation of a periodicity which is also about 1.5 times as long.

\begin{figure}[tb]
	\includegraphics[width=0.9\columnwidth]{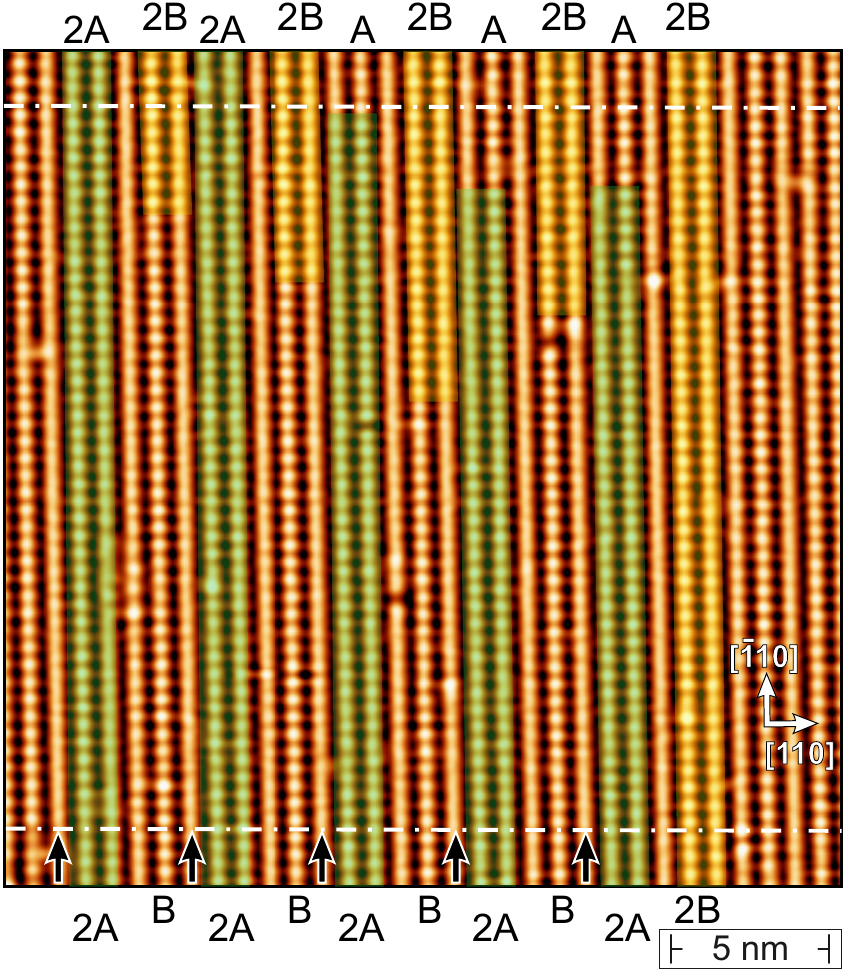}
			\caption{Large SP-STM overview scan performed with a magnetic tip on MnO$_2$ chains on Pt(001). 
					The intra-chain coupling is not strictly collinear, 
					as can be recognized by inspecting the MnO$_2$ chains marked by black arrows.  
					Whereas no significant magnetic contrast is detected in the bottom part, 
					a strong magnetic contrast with a $2a_\textrm{Pt}$ periodicity 
					is clearly visible in the upper part of the image. }
\label{Fig:MnO-Pt-large}
\end{figure}
As we will show below deviations from the $(15\,\times\,2)$ magnetic unit cell 
can be detected by imaging MnO$_2$ chains on Pt(001) over larger surface areas.
For example, Fig.\,\ref{Fig:MnO-Pt-large} shows an SP-STM image with a scan area of 30\,nm\,$\times$\,30\,nm. 
For better visibility, the data set was filtered by using a band pass. 
As described above SP-STM images of the $(15\,\times\,2)$ MnO$_2$/Pt(001) magnetic unit cell 
are characterized by five-fold periodicity of the structural $(3\,\times\,1)$ unit cell.  
In general, the magnetization direction-dependent variation of the total current gives rise to 
corrugation maxima if tip and sample are magnetized in the same direction, 
corrugation minima for an antiparrallel alignment, or a very low or even vanishing corrugation 
if the angle between the tip and sample magnetization direction is close to 90$^{\circ}$ (cf.\ Eq.\,\ref{Eq:SP-STM}).  
At the $y$-position marked by a dash-dotted white line at the bottom of Fig.\,\ref{Fig:MnO-Pt-large} 
we find a sequence consisting of two MnO$_2$ chains which exhibit corrugation maxima (colored green, labelled 2A), 
then one chain with a low contrast, one MnO$_2$ chains with corrugation minima (B), 
and finally a chain with no recognizable spin contrast (marked by black arrows). 
Close inspection of the data set presented in Fig.\,\ref{Fig:MnO-Pt-large} reveals 
that the contrast of the specific MnO$_2$ chains is not constant but slowly varies.  
For example, the MnO$_2$ chains marked by black arrows which showed a vanishingly low magnetic contrast 
at the bottom of Fig.\,\ref{Fig:MnO-Pt-large} develop a sizable magnetic contrast towards the top part of the scan.  
This observation indicates that the magnetic order along the MnO$_2$ chains is not strictly collinear but slightly canted. 
Within our field of view of 30\,nm along the $[{\bar 1}10]$ direction, however, we only observe about a quarter of a $2\pi$ rotation. 
Therefore, we can conclude that the wavelength of the magnetic structure 
along the MnO$_2$ chains on Pt(001) must be very long, probably of the order of 100\,nm. 

As a result of the spin rotation along the MnO$_2$ chains and the existence 
of occasional structural defects the magnetic order is not strictly periodic.  
In fact, comparison of the bottom part of the SP-STM image 
with the top part of Fig.\,\ref{Fig:MnO-Pt-large} reveals that the contrast changed. 
Whereas the sequence of contrasts obtained within the five TMO chains of a magnetic unit cell 
was 2A-low-B-low in the bottom part, the dominating sequence in the upper part is A-low-2B-low.  
At the moment we can only speculate about the origin of this non-collinearity along the chains.  
Possibly, DM-type interactions triggered by the overlap of the O $2p$ with Ir $5d$ orbitals 
also influence the superexchange along the TMO chains.  

\section{Conclusion}
In summary, we systematically investigated the structural and magnetic properties 
of transition metal oxides (TMOs) on the fcc(001) surfaces of Ir and Pt. 
We find that Co, Fe, and Mn form quasi one-dimensional TMO chains with a $(3\,\times\,1)$ unit cell. 
Similar $(3\,\times\,1)$-ordered chains were observed for Cr on Ir(001). 
Whereas no magnetic signal was found for Co- and Cr-based chains, our SP-STM measurements confirm 
the theoretically predicted antiferromagnetic intra-chain coupling 
for FeO$_2$/Ir(001) and for MnO$_2$ chains on both substrates. 
A ferromagnetic inter-chain coupling is found for FeO$_2$/Ir(001). 
In the case of MnO$_2$ the SP-STM images reveal a complex helical intra-chain magnetic coupling, 
resulting in a $(9\,\times\,2)$ magnetic unit cell on Ir(001) and a $(15\,\times\,2)$ magnetic unit cell for Pt(001).  
Furthermore, large scale SP-STM images of MnO$_2$ chains on Pt(001) unveil 
that also the intra-chain magnetic coupling is not perfectly collinear but slightly canted, 
resulting in a spin spiral with a periodicity corresponding to several hundreds of substrate atoms.  
Our results highlight the relevance of spin-orbit--related effects for magnetic coupling phenomena 
in systems with broken inversion symmetry, such as surfaces or nanoparticles.  
\vspace{0.0cm}

\section*{Acknowledgments} 

Experimental work was supported by DFG through FOR 1700 (project E6), SPP 2137 ``Skyrmionics'' (BO 1468/26-1) 
and by the Dresden-W{\"u}rzburg Center for Topological Quantum Matter Research (ct.qmat)..
The authors would like to thank Paolo Moras (Trieste, Italy) for bringing one-dimensional TMOs to our attention 
and Jens K{\"u}gel (W{\"u}rzburg, Germany) for critically reading the manuscript.


%


\begin{thebibliography}{37}%
\makeatletter
\providecommand \@ifxundefined [1]{%
 \@ifx{#1\undefined}
}%
\providecommand \@ifnum [1]{%
 \ifnum #1\expandafter \@firstoftwo
 \else \expandafter \@secondoftwo
 \fi
}%
\providecommand \@ifx [1]{%
 \ifx #1\expandafter \@firstoftwo
 \else \expandafter \@secondoftwo
 \fi
}%
\providecommand \natexlab [1]{#1}%
\providecommand \enquote  [1]{``#1''}%
\providecommand \bibnamefont  [1]{#1}%
\providecommand \bibfnamefont [1]{#1}%
\providecommand \citenamefont [1]{#1}%
\providecommand \href@noop [0]{\@secondoftwo}%
\providecommand \href [0]{\begingroup \@sanitize@url \@href}%
\providecommand \@href[1]{\@@startlink{#1}\@@href}%
\providecommand \@@href[1]{\endgroup#1\@@endlink}%
\providecommand \@sanitize@url [0]{\catcode `\\12\catcode `\$12\catcode
  `\&12\catcode `\#12\catcode `\^12\catcode `\_12\catcode `\%12\relax}%
\providecommand \@@startlink[1]{}%
\providecommand \@@endlink[0]{}%
\providecommand \url  [0]{\begingroup\@sanitize@url \@url }%
\providecommand \@url [1]{\endgroup\@href {#1}{\urlprefix }}%
\providecommand \urlprefix  [0]{URL }%
\providecommand \Eprint [0]{\href }%
\providecommand \doibase [0]{http://dx.doi.org/}%
\providecommand \selectlanguage [0]{\@gobble}%
\providecommand \bibinfo  [0]{\@secondoftwo}%
\providecommand \bibfield  [0]{\@secondoftwo}%
\providecommand \translation [1]{[#1]}%
\providecommand \BibitemOpen [0]{}%
\providecommand \bibitemStop [0]{}%
\providecommand \bibitemNoStop [0]{.\EOS\space}%
\providecommand \EOS [0]{\spacefactor3000\relax}%
\providecommand \BibitemShut  [1]{\csname bibitem#1\endcsname}%
\let\auto@bib@innerbib\@empty
\bibitem [{\citenamefont {Vaz}\ \emph {et~al.}(2008)\citenamefont {Vaz},
  \citenamefont {Bland},\ and\ \citenamefont {Lauhoff}}]{Vaz2008}%
  \BibitemOpen
  \bibfield  {author} {\bibinfo {author} {\bibfnamefont {C.~A.~F.}\
  \bibnamefont {Vaz}}, \bibinfo {author} {\bibfnamefont {J.~A.~C.}\
  \bibnamefont {Bland}}, \ and\ \bibinfo {author} {\bibfnamefont
  {G.}~\bibnamefont {Lauhoff}},\ }\bibfield  {title} {\enquote {\bibinfo
  {title} {Magnetism in ultrathin film structures},}\ }\href {\doibase
  10.1088/0034-4885/71/5/056501} {\bibfield  {journal} {\bibinfo  {journal}
  {Reports on Progress in Physics}\ }\textbf {\bibinfo {volume} {71}},\
  \bibinfo {pages} {056501} (\bibinfo {year} {2008})}\BibitemShut {NoStop}%
\bibitem [{\citenamefont {Braun}(2012)}]{Braun2012}%
  \BibitemOpen
  \bibfield  {author} {\bibinfo {author} {\bibfnamefont {H.-B.}\ \bibnamefont
  {Braun}},\ }\bibfield  {title} {\enquote {\bibinfo {title} {Topological
  effects in nanomagnetism: from superparamagnetism to chiral quantum
  solitons},}\ }\href {\doibase 10.1080/00018732.2012.663070} {\bibfield
  {journal} {\bibinfo  {journal} {Advances in Physics}\ }\textbf {\bibinfo
  {volume} {61}},\ \bibinfo {pages} {1} (\bibinfo {year} {2012})}\BibitemShut
  {NoStop}%
\bibitem [{\citenamefont {Bode}(2003)}]{Bode2003}%
  \BibitemOpen
  \bibfield  {author} {\bibinfo {author} {\bibfnamefont {M.}~\bibnamefont
  {Bode}},\ }\bibfield  {title} {\enquote {\bibinfo {title} {Spin-polarized
  scanning tunnelling microscopy},}\ }\href {\doibase
  10.1088/0034-4885/66/4/203} {\bibfield  {journal} {\bibinfo  {journal}
  {Reports on Progress in Physics}\ }\textbf {\bibinfo {volume} {66}},\
  \bibinfo {pages} {523} (\bibinfo {year} {2003})}\BibitemShut {NoStop}%
\bibitem [{\citenamefont {Kubetzka}\ \emph {et~al.}(2005)\citenamefont
  {Kubetzka}, \citenamefont {Ferriani}, \citenamefont {Bode}, \citenamefont
  {Heinze}, \citenamefont {Bihlmayer}, \citenamefont {von Bergmann},
  \citenamefont {Pietzsch}, \citenamefont {Bl\"ugel},\ and\ \citenamefont
  {Wiesendanger}}]{KFB2005}%
  \BibitemOpen
  \bibfield  {author} {\bibinfo {author} {\bibfnamefont {A.}~\bibnamefont
  {Kubetzka}}, \bibinfo {author} {\bibfnamefont {P.}~\bibnamefont {Ferriani}},
  \bibinfo {author} {\bibfnamefont {M.}~\bibnamefont {Bode}}, \bibinfo {author}
  {\bibfnamefont {S.}~\bibnamefont {Heinze}}, \bibinfo {author} {\bibfnamefont
  {G.}~\bibnamefont {Bihlmayer}}, \bibinfo {author} {\bibfnamefont
  {K.}~\bibnamefont {von Bergmann}}, \bibinfo {author} {\bibfnamefont
  {O.}~\bibnamefont {Pietzsch}}, \bibinfo {author} {\bibfnamefont
  {S.}~\bibnamefont {Bl\"ugel}}, \ and\ \bibinfo {author} {\bibfnamefont
  {R.}~\bibnamefont {Wiesendanger}},\ }\bibfield  {title} {\enquote {\bibinfo
  {title} {Revealing antiferromagnetic order of the {F}e monolayer on {W}(001):
  Spin-polarized scanning tunneling microscopy and first-principles
  calculations},}\ }\href {\doibase 10.1103/PhysRevLett.94.087204} {\bibfield
  {journal} {\bibinfo  {journal} {Phys. Rev. Lett.}\ }\textbf {\bibinfo
  {volume} {94}},\ \bibinfo {pages} {087204} (\bibinfo {year}
  {2005})}\BibitemShut {NoStop}%
\bibitem [{\citenamefont {Bode}\ \emph {et~al.}(2006)\citenamefont {Bode},
  \citenamefont {Vedmedenko}, \citenamefont {von Bergmann}, \citenamefont
  {Kubetzka}, \citenamefont {Ferriani}, \citenamefont {Heinze},\ and\
  \citenamefont {Wiesendanger}}]{BVB2006}%
  \BibitemOpen
  \bibfield  {author} {\bibinfo {author} {\bibfnamefont {M.}~\bibnamefont
  {Bode}}, \bibinfo {author} {\bibfnamefont {E.~Y.}\ \bibnamefont
  {Vedmedenko}}, \bibinfo {author} {\bibfnamefont {K.}~\bibnamefont {von
  Bergmann}}, \bibinfo {author} {\bibfnamefont {A.}~\bibnamefont {Kubetzka}},
  \bibinfo {author} {\bibfnamefont {P.}~\bibnamefont {Ferriani}}, \bibinfo
  {author} {\bibfnamefont {S.}~\bibnamefont {Heinze}}, \ and\ \bibinfo {author}
  {\bibfnamefont {R.}~\bibnamefont {Wiesendanger}},\ }\bibfield  {title}
  {\enquote {\bibinfo {title} {Atomic spin structure of antiferromagnetic
  domain walls},}\ }\href {\doibase 10.1038/nmat1646} {\bibfield  {journal}
  {\bibinfo  {journal} {Nature Materials}\ }\textbf {\bibinfo {volume} {5}},\
  \bibinfo {pages} {477} (\bibinfo {year} {2006})}\BibitemShut {NoStop}%
\bibitem [{\citenamefont {Gao}\ \emph {et~al.}(2008)\citenamefont {Gao},
  \citenamefont {Wulfhekel},\ and\ \citenamefont {Kirschner}}]{GWK2008}%
  \BibitemOpen
  \bibfield  {author} {\bibinfo {author} {\bibfnamefont {C.~L.}\ \bibnamefont
  {Gao}}, \bibinfo {author} {\bibfnamefont {W.}~\bibnamefont {Wulfhekel}}, \
  and\ \bibinfo {author} {\bibfnamefont {J.}~\bibnamefont {Kirschner}},\
  }\bibfield  {title} {\enquote {\bibinfo {title} {Revealing the $120^{\circ}$
  antiferromagnetic {N{\'{e}}el} structure in real space: One monolayer {Mn} on
  {Ag(111)}},}\ }\href
  {https://link.aps.org/doi/10.1103/PhysRevLett.101.267205} {\bibfield
  {journal} {\bibinfo  {journal} {Phys. Rev. Lett.}\ }\textbf {\bibinfo
  {volume} {101}},\ \bibinfo {pages} {267205} (\bibinfo {year}
  {2008})}\BibitemShut {NoStop}%
\bibitem [{\citenamefont {Bode}\ \emph {et~al.}(2002)\citenamefont {Bode},
  \citenamefont {Heinze}, \citenamefont {Kubetzka}, \citenamefont {Pietzsch},
  \citenamefont {Hennefarth}, \citenamefont {Getzlaff}, \citenamefont
  {Wiesendanger}, \citenamefont {Nie}, \citenamefont {Bihlmayer},\ and\
  \citenamefont {Bl\"ugel}}]{PhysRevB.66.014425}%
  \BibitemOpen
  \bibfield  {author} {\bibinfo {author} {\bibfnamefont {M.}~\bibnamefont
  {Bode}}, \bibinfo {author} {\bibfnamefont {S.}~\bibnamefont {Heinze}},
  \bibinfo {author} {\bibfnamefont {A.}~\bibnamefont {Kubetzka}}, \bibinfo
  {author} {\bibfnamefont {O.}~\bibnamefont {Pietzsch}}, \bibinfo {author}
  {\bibfnamefont {M.}~\bibnamefont {Hennefarth}}, \bibinfo {author}
  {\bibfnamefont {M.}~\bibnamefont {Getzlaff}}, \bibinfo {author}
  {\bibfnamefont {R.}~\bibnamefont {Wiesendanger}}, \bibinfo {author}
  {\bibfnamefont {X.}~\bibnamefont {Nie}}, \bibinfo {author} {\bibfnamefont
  {G.}~\bibnamefont {Bihlmayer}}, \ and\ \bibinfo {author} {\bibfnamefont
  {S.}~\bibnamefont {Bl\"ugel}},\ }\bibfield  {title} {\enquote {\bibinfo
  {title} {Structural, electronic, and magnetic properties of a {M}n monolayer
  on {W}(110)},}\ }\href {\doibase 10.1103/PhysRevB.66.014425} {\bibfield
  {journal} {\bibinfo  {journal} {Phys. Rev. B}\ }\textbf {\bibinfo {volume}
  {66}},\ \bibinfo {pages} {014425} (\bibinfo {year} {2002})}\BibitemShut
  {NoStop}%
\bibitem [{\citenamefont {Heinze}\ \emph {et~al.}(2000)\citenamefont {Heinze},
  \citenamefont {Bode}, \citenamefont {Kubetzka}, \citenamefont {Pietzsch},
  \citenamefont {Nie}, \citenamefont {Bl{\"u}gel},\ and\ \citenamefont
  {Wiesendanger}}]{Heinze2000}%
  \BibitemOpen
  \bibfield  {author} {\bibinfo {author} {\bibfnamefont {S.}~\bibnamefont
  {Heinze}}, \bibinfo {author} {\bibfnamefont {M.}~\bibnamefont {Bode}},
  \bibinfo {author} {\bibfnamefont {A.}~\bibnamefont {Kubetzka}}, \bibinfo
  {author} {\bibfnamefont {O.}~\bibnamefont {Pietzsch}}, \bibinfo {author}
  {\bibfnamefont {X.}~\bibnamefont {Nie}}, \bibinfo {author} {\bibfnamefont
  {S.}~\bibnamefont {Bl{\"u}gel}}, \ and\ \bibinfo {author} {\bibfnamefont
  {R.}~\bibnamefont {Wiesendanger}},\ }\bibfield  {title} {\enquote {\bibinfo
  {title} {Real-space imaging of two-dimensional antiferromagnetism on the
  atomic scale},}\ }\href {\doibase 10.1126/science.288.5472.1805} {\bibfield
  {journal} {\bibinfo  {journal} {Science}\ }\textbf {\bibinfo {volume}
  {288}},\ \bibinfo {pages} {1805} (\bibinfo {year} {2000})}\BibitemShut
  {NoStop}%
\bibitem [{\citenamefont {Bode}\ \emph {et~al.}(2007)\citenamefont {Bode},
  \citenamefont {Heide}, \citenamefont {von Bergmann}, \citenamefont
  {Ferriani}, \citenamefont {Heinze}, \citenamefont {Bihlmayer}, \citenamefont
  {Kubetzka}, \citenamefont {Pietzsch}, \citenamefont {Bl{\"u}gel},\ and\
  \citenamefont {Wiesendanger}}]{Bode2007}%
  \BibitemOpen
  \bibfield  {author} {\bibinfo {author} {\bibfnamefont {M.}~\bibnamefont
  {Bode}}, \bibinfo {author} {\bibfnamefont {M.}~\bibnamefont {Heide}},
  \bibinfo {author} {\bibfnamefont {K.}~\bibnamefont {von Bergmann}}, \bibinfo
  {author} {\bibfnamefont {P.}~\bibnamefont {Ferriani}}, \bibinfo {author}
  {\bibfnamefont {S.}~\bibnamefont {Heinze}}, \bibinfo {author} {\bibfnamefont
  {G.}~\bibnamefont {Bihlmayer}}, \bibinfo {author} {\bibfnamefont
  {A.}~\bibnamefont {Kubetzka}}, \bibinfo {author} {\bibfnamefont
  {O.}~\bibnamefont {Pietzsch}}, \bibinfo {author} {\bibfnamefont
  {S.}~\bibnamefont {Bl{\"u}gel}}, \ and\ \bibinfo {author} {\bibfnamefont
  {R.}~\bibnamefont {Wiesendanger}},\ }\bibfield  {title} {\enquote {\bibinfo
  {title} {Chiral magnetic order at surfaces driven by inversion asymmetry},}\
  }\href {https://doi.org/10.1038/nature05802} {\bibfield  {journal} {\bibinfo
  {journal} {Nature}\ }\textbf {\bibinfo {volume} {447}},\ \bibinfo {pages}
  {190} (\bibinfo {year} {2007})}\BibitemShut {NoStop}%
\bibitem [{\citenamefont {Ferstl}\ \emph {et~al.}(2016)\citenamefont {Ferstl},
  \citenamefont {Hammer}, \citenamefont {Sobel}, \citenamefont {Gubo},
  \citenamefont {Heinz}, \citenamefont {Schneider}, \citenamefont
  {Mittendorfer},\ and\ \citenamefont {Redinger}}]{Ferstl2016}%
  \BibitemOpen
  \bibfield  {author} {\bibinfo {author} {\bibfnamefont {P.}~\bibnamefont
  {Ferstl}}, \bibinfo {author} {\bibfnamefont {L.}~\bibnamefont {Hammer}},
  \bibinfo {author} {\bibfnamefont {C.}~\bibnamefont {Sobel}}, \bibinfo
  {author} {\bibfnamefont {M.}~\bibnamefont {Gubo}}, \bibinfo {author}
  {\bibfnamefont {K.}~\bibnamefont {Heinz}}, \bibinfo {author} {\bibfnamefont
  {M.~A.}\ \bibnamefont {Schneider}}, \bibinfo {author} {\bibfnamefont
  {F.}~\bibnamefont {Mittendorfer}}, \ and\ \bibinfo {author} {\bibfnamefont
  {J.}~\bibnamefont {Redinger}},\ }\bibfield  {title} {\enquote {\bibinfo
  {title} {Self-organized growth, structure, and magnetism of monatomic
  transition-metal oxide chains},}\ }\href {\doibase
  10.1103/PhysRevLett.117.046101} {\bibfield  {journal} {\bibinfo  {journal}
  {Phys. Rev. Lett.}\ }\textbf {\bibinfo {volume} {117}},\ \bibinfo {pages}
  {046101} (\bibinfo {year} {2016})}\BibitemShut {NoStop}%
\bibitem [{\citenamefont {Ferstl}\ \emph {et~al.}(2017)\citenamefont {Ferstl},
  \citenamefont {Mittendorfer}, \citenamefont {Redinger}, \citenamefont
  {Schneider},\ and\ \citenamefont {Hammer}}]{Ferstl2017}%
  \BibitemOpen
  \bibfield  {author} {\bibinfo {author} {\bibfnamefont {P.}~\bibnamefont
  {Ferstl}}, \bibinfo {author} {\bibfnamefont {F.}~\bibnamefont
  {Mittendorfer}}, \bibinfo {author} {\bibfnamefont {J.}~\bibnamefont
  {Redinger}}, \bibinfo {author} {\bibfnamefont {M.~A.}\ \bibnamefont
  {Schneider}}, \ and\ \bibinfo {author} {\bibfnamefont {L.}~\bibnamefont
  {Hammer}},\ }\bibfield  {title} {\enquote {\bibinfo {title} {Monatomic {Co},
  {C}o{O}$_{2}$, and {C}o{O}$_{3}$ nanowires on {I}r(100) and {P}t(100)
  surfaces: Formation, structure, and energetics},}\ }\href {\doibase
  10.1103/PhysRevB.96.085407} {\bibfield  {journal} {\bibinfo  {journal} {Phys.
  Rev. B}\ }\textbf {\bibinfo {volume} {96}},\ \bibinfo {pages} {085407}
  (\bibinfo {year} {2017})}\BibitemShut {NoStop}%
\bibitem [{\citenamefont {Kr\"onlein}\ \emph {et~al.}(2018)\citenamefont
  {Kr\"onlein}, \citenamefont {Schmitt}, \citenamefont {Hoffmann},
  \citenamefont {Kemmer}, \citenamefont {Seubert}, \citenamefont {Vogt},
  \citenamefont {K\"uspert}, \citenamefont {B\"ohme}, \citenamefont {Alonazi},
  \citenamefont {K\"ugel}, \citenamefont {Albrithen}, \citenamefont {Bode},
  \citenamefont {Bihlmayer},\ and\ \citenamefont {Bl\"ugel}}]{KSH2018}%
  \BibitemOpen
  \bibfield  {author} {\bibinfo {author} {\bibfnamefont {A.}~\bibnamefont
  {Kr\"onlein}}, \bibinfo {author} {\bibfnamefont {M.}~\bibnamefont {Schmitt}},
  \bibinfo {author} {\bibfnamefont {M.}~\bibnamefont {Hoffmann}}, \bibinfo
  {author} {\bibfnamefont {J.}~\bibnamefont {Kemmer}}, \bibinfo {author}
  {\bibfnamefont {N.}~\bibnamefont {Seubert}}, \bibinfo {author} {\bibfnamefont
  {M.}~\bibnamefont {Vogt}}, \bibinfo {author} {\bibfnamefont {J.}~\bibnamefont
  {K\"uspert}}, \bibinfo {author} {\bibfnamefont {M.}~\bibnamefont {B\"ohme}},
  \bibinfo {author} {\bibfnamefont {B.}~\bibnamefont {Alonazi}}, \bibinfo
  {author} {\bibfnamefont {J.}~\bibnamefont {K\"ugel}}, \bibinfo {author}
  {\bibfnamefont {H.~A.}\ \bibnamefont {Albrithen}}, \bibinfo {author}
  {\bibfnamefont {M.}~\bibnamefont {Bode}}, \bibinfo {author} {\bibfnamefont
  {G.}~\bibnamefont {Bihlmayer}}, \ and\ \bibinfo {author} {\bibfnamefont
  {S.}~\bibnamefont {Bl\"ugel}},\ }\bibfield  {title} {\enquote {\bibinfo
  {title} {Magnetic ground state stabilized by three-site interactions:
  $\mathrm{Fe}/\mathrm{Rh}(111)$},}\ }\href {\doibase
  10.1103/PhysRevLett.120.207202} {\bibfield  {journal} {\bibinfo  {journal}
  {Phys. Rev. Lett.}\ }\textbf {\bibinfo {volume} {120}},\ \bibinfo {pages}
  {207202} (\bibinfo {year} {2018})}\BibitemShut {NoStop}%
\bibitem [{\citenamefont {Romming}\ \emph {et~al.}(2018)\citenamefont
  {Romming}, \citenamefont {Pralow}, \citenamefont {Kubetzka}, \citenamefont
  {Hoffmann}, \citenamefont {von Malottki}, \citenamefont {Meyer},
  \citenamefont {Dup{\'{e}}}, \citenamefont {Wiesendanger}, \citenamefont {von
  Bergmann},\ and\ \citenamefont {Heinze}}]{RPK2018}%
  \BibitemOpen
  \bibfield  {author} {\bibinfo {author} {\bibfnamefont {N.}~\bibnamefont
  {Romming}}, \bibinfo {author} {\bibfnamefont {H.}~\bibnamefont {Pralow}},
  \bibinfo {author} {\bibfnamefont {A.}~\bibnamefont {Kubetzka}}, \bibinfo
  {author} {\bibfnamefont {M.}~\bibnamefont {Hoffmann}}, \bibinfo {author}
  {\bibfnamefont {S.}~\bibnamefont {von Malottki}}, \bibinfo {author}
  {\bibfnamefont {S.}~\bibnamefont {Meyer}}, \bibinfo {author} {\bibfnamefont
  {B.}~\bibnamefont {Dup{\'{e}}}}, \bibinfo {author} {\bibfnamefont
  {R.}~\bibnamefont {Wiesendanger}}, \bibinfo {author} {\bibfnamefont
  {K.}~\bibnamefont {von Bergmann}}, \ and\ \bibinfo {author} {\bibfnamefont
  {S.}~\bibnamefont {Heinze}},\ }\bibfield  {title} {\enquote {\bibinfo {title}
  {Competition of {D}zyaloshinskii-{M}oriya and higher-order exchange
  interactions in {Rh}/{Fe} atomic bilayers on {I}r(111)},}\ }\href {\doibase
  10.1103/PhysRevLett.120.207201} {\bibfield  {journal} {\bibinfo  {journal}
  {Phys. Rev. Lett.}\ }\textbf {\bibinfo {volume} {120}},\ \bibinfo {pages}
  {207201} (\bibinfo {year} {2018})}\BibitemShut {NoStop}%
\bibitem [{\citenamefont {Schmitt}\ \emph {et~al.}(2019)\citenamefont
  {Schmitt}, \citenamefont {Moras}, \citenamefont {Bihlmayer}, \citenamefont
  {Cotsakis}, \citenamefont {Vogt}, \citenamefont {Kemmer}, \citenamefont
  {Belabbes}, \citenamefont {Sheverdyaeva}, \citenamefont {Kundu},
  \citenamefont {Carbone}, \citenamefont {Bl{\"u}gel},\ and\ \citenamefont
  {Bode}}]{NC}%
  \BibitemOpen
  \bibfield  {author} {\bibinfo {author} {\bibfnamefont {M.}~\bibnamefont
  {Schmitt}}, \bibinfo {author} {\bibfnamefont {P.}~\bibnamefont {Moras}},
  \bibinfo {author} {\bibfnamefont {G.}~\bibnamefont {Bihlmayer}}, \bibinfo
  {author} {\bibfnamefont {R.}~\bibnamefont {Cotsakis}}, \bibinfo {author}
  {\bibfnamefont {M.}~\bibnamefont {Vogt}}, \bibinfo {author} {\bibfnamefont
  {J.}~\bibnamefont {Kemmer}}, \bibinfo {author} {\bibfnamefont
  {A.}~\bibnamefont {Belabbes}}, \bibinfo {author} {\bibfnamefont {P.~M.}\
  \bibnamefont {Sheverdyaeva}}, \bibinfo {author} {\bibfnamefont {A.~K.}\
  \bibnamefont {Kundu}}, \bibinfo {author} {\bibfnamefont {C.}~\bibnamefont
  {Carbone}}, \bibinfo {author} {\bibfnamefont {S.}~\bibnamefont {Bl{\"u}gel}},
  \ and\ \bibinfo {author} {\bibfnamefont {M.}~\bibnamefont {Bode}},\
  }\bibfield  {title} {\enquote {\bibinfo {title} {Indirect chiral magnetic
  exchange through {D}zyaloshinskii-{M}oriya-enhanced {RKKY} interactions in
  manganese oxide chains on {I}r(100)},}\ }\href {\doibase
  10.1038/s41467-019-10515-3} {\bibfield  {journal} {\bibinfo  {journal}
  {Nature Communications}\ }\textbf {\bibinfo {volume} {10}},\ \bibinfo {pages}
  {2610} (\bibinfo {year} {2019})}\BibitemShut {NoStop}%
\bibitem [{\citenamefont {Khajetoorians}\ \emph {et~al.}(2016)\citenamefont
  {Khajetoorians}, \citenamefont {Steinbrecher}, \citenamefont {Ternes},
  \citenamefont {Bouhassoune}, \citenamefont {dos Santos~Dias}, \citenamefont
  {Lounis}, \citenamefont {Wiebe},\ and\ \citenamefont
  {Wiesendanger}}]{Khajetoorians2016}%
  \BibitemOpen
  \bibfield  {author} {\bibinfo {author} {\bibfnamefont {A.~A.}\ \bibnamefont
  {Khajetoorians}}, \bibinfo {author} {\bibfnamefont {M.}~\bibnamefont
  {Steinbrecher}}, \bibinfo {author} {\bibfnamefont {M.}~\bibnamefont
  {Ternes}}, \bibinfo {author} {\bibfnamefont {M.}~\bibnamefont {Bouhassoune}},
  \bibinfo {author} {\bibfnamefont {M.}~\bibnamefont {dos Santos~Dias}},
  \bibinfo {author} {\bibfnamefont {S.}~\bibnamefont {Lounis}}, \bibinfo
  {author} {\bibfnamefont {J.}~\bibnamefont {Wiebe}}, \ and\ \bibinfo {author}
  {\bibfnamefont {R.}~\bibnamefont {Wiesendanger}},\ }\bibfield  {title}
  {\enquote {\bibinfo {title} {Tailoring the chiral magnetic interaction
  between two individual atoms},}\ }\href {\doibase 10.1038/ncomms10620}
  {\bibfield  {journal} {\bibinfo  {journal} {Nature Communications}\ }\textbf
  {\bibinfo {volume} {7}},\ \bibinfo {pages} {10620} (\bibinfo {year}
  {2016})}\BibitemShut {NoStop}%
\bibitem [{\citenamefont {Hermenau}\ \emph {et~al.}(2017)\citenamefont
  {Hermenau}, \citenamefont {Iba{\~n}ez-Azpiroz}, \citenamefont {H{\"u}bner},
  \citenamefont {Sonntag}, \citenamefont {Baxevanis}, \citenamefont {Ton},
  \citenamefont {Steinbrecher}, \citenamefont {Khajetoorians}, \citenamefont
  {dos Santos~Dias}, \citenamefont {Bl{\"u}gel}, \citenamefont {Wiesendanger},
  \citenamefont {Lounis},\ and\ \citenamefont {Wiebe}}]{Hermenau2017}%
  \BibitemOpen
  \bibfield  {author} {\bibinfo {author} {\bibfnamefont {J.}~\bibnamefont
  {Hermenau}}, \bibinfo {author} {\bibfnamefont {J.}~\bibnamefont
  {Iba{\~n}ez-Azpiroz}}, \bibinfo {author} {\bibfnamefont {C.}~\bibnamefont
  {H{\"u}bner}}, \bibinfo {author} {\bibfnamefont {A.}~\bibnamefont {Sonntag}},
  \bibinfo {author} {\bibfnamefont {B.}~\bibnamefont {Baxevanis}}, \bibinfo
  {author} {\bibfnamefont {K.~T.}\ \bibnamefont {Ton}}, \bibinfo {author}
  {\bibfnamefont {M.}~\bibnamefont {Steinbrecher}}, \bibinfo {author}
  {\bibfnamefont {A.~A.}\ \bibnamefont {Khajetoorians}}, \bibinfo {author}
  {\bibfnamefont {M.}~\bibnamefont {dos Santos~Dias}}, \bibinfo {author}
  {\bibfnamefont {S.}~\bibnamefont {Bl{\"u}gel}}, \bibinfo {author}
  {\bibfnamefont {R.}~\bibnamefont {Wiesendanger}}, \bibinfo {author}
  {\bibfnamefont {S.}~\bibnamefont {Lounis}}, \ and\ \bibinfo {author}
  {\bibfnamefont {J.}~\bibnamefont {Wiebe}},\ }\bibfield  {title} {\enquote
  {\bibinfo {title} {A gateway towards non-collinear spin processing using
  three-atom magnets with strong substrate coupling},}\ }\href
  {https://doi.org/10.1038/s41467-017-00506-7} {\bibfield  {journal} {\bibinfo
  {journal} {Nature Communications}\ }\textbf {\bibinfo {volume} {8}},\
  \bibinfo {pages} {642} (\bibinfo {year} {2017})}\BibitemShut {NoStop}%
\bibitem [{\citenamefont {Grant}(1969)}]{Grant1969}%
  \BibitemOpen
  \bibfield  {author} {\bibinfo {author} {\bibfnamefont {J.~T.}\ \bibnamefont
  {Grant}},\ }\bibfield  {title} {\enquote {\bibinfo {title} {A {LEED} study of
  the {I}r(100) surface},}\ }\href
  {http://www.sciencedirect.com/science/article/pii/0039602869901678}
  {\bibfield  {journal} {\bibinfo  {journal} {Surface Science}\ }\textbf
  {\bibinfo {volume} {18}},\ \bibinfo {pages} {228} (\bibinfo {year}
  {1969})}\BibitemShut {NoStop}%
\bibitem [{\citenamefont {Rhodin}\ and\ \citenamefont
  {Brod{\'{e}}n}(1976)}]{Rhodin1976}%
  \BibitemOpen
  \bibfield  {author} {\bibinfo {author} {\bibfnamefont {T.~N.}\ \bibnamefont
  {Rhodin}}\ and\ \bibinfo {author} {\bibfnamefont {G.}~\bibnamefont
  {Brod{\'{e}}n}},\ }\bibfield  {title} {\enquote {\bibinfo {title}
  {Preparation and chemisorptive properties of the clean normal and
  reconstructed surfaces of {I}r(100): role of multiplets},}\ }\href {\doibase
  10.1016/0039-6028(76)90329-0} {\bibfield  {journal} {\bibinfo  {journal}
  {Surface Science}\ }\textbf {\bibinfo {volume} {60}},\ \bibinfo {pages} {466}
  (\bibinfo {year} {1976})}\BibitemShut {NoStop}%
\bibitem [{\citenamefont {Schmidt}\ \emph {et~al.}(2002)\citenamefont
  {Schmidt}, \citenamefont {Meier}, \citenamefont {Hammer},\ and\ \citenamefont
  {Heinz}}]{Schmidt2002}%
  \BibitemOpen
  \bibfield  {author} {\bibinfo {author} {\bibfnamefont {A.}~\bibnamefont
  {Schmidt}}, \bibinfo {author} {\bibfnamefont {W.}~\bibnamefont {Meier}},
  \bibinfo {author} {\bibfnamefont {L.}~\bibnamefont {Hammer}}, \ and\ \bibinfo
  {author} {\bibfnamefont {K.}~\bibnamefont {Heinz}},\ }\bibfield  {title}
  {\enquote {\bibinfo {title} {Deep-going reconstruction of
  {I}r(100)-$5\,\times\,1$},}\ }\href {\doibase 10.1088/0953-8984/14/47/310}
  {\bibfield  {journal} {\bibinfo  {journal} {Journal of Physics: Condensed
  Matter}\ }\textbf {\bibinfo {volume} {14}},\ \bibinfo {pages} {12353}
  (\bibinfo {year} {2002})}\BibitemShut {NoStop}%
\bibitem [{\citenamefont {Hammer}\ \emph
  {et~al.}(2016{\natexlab{a}})\citenamefont {Hammer}, \citenamefont {Meinel},
  \citenamefont {Krahn},\ and\ \citenamefont {Widdra}}]{Hammer2016}%
  \BibitemOpen
  \bibfield  {author} {\bibinfo {author} {\bibfnamefont {R.}~\bibnamefont
  {Hammer}}, \bibinfo {author} {\bibfnamefont {K.}~\bibnamefont {Meinel}},
  \bibinfo {author} {\bibfnamefont {O.}~\bibnamefont {Krahn}}, \ and\ \bibinfo
  {author} {\bibfnamefont {W.}~\bibnamefont {Widdra}},\ }\bibfield  {title}
  {\enquote {\bibinfo {title} {Surface reconstruction of {P}t(001)
  quantitatively revisited},}\ }\href {\doibase 10.1103/PhysRevB.94.195406}
  {\bibfield  {journal} {\bibinfo  {journal} {Phys. Rev. B}\ }\textbf {\bibinfo
  {volume} {94}},\ \bibinfo {pages} {195406} (\bibinfo {year}
  {2016}{\natexlab{a}})}\BibitemShut {NoStop}%
\bibitem [{\citenamefont {Morgan}\ and\ \citenamefont
  {Somorjai}(1968)}]{Morgan1968}%
  \BibitemOpen
  \bibfield  {author} {\bibinfo {author} {\bibfnamefont {A.~E.}\ \bibnamefont
  {Morgan}}\ and\ \bibinfo {author} {\bibfnamefont {G.~A.}\ \bibnamefont
  {Somorjai}},\ }\bibfield  {title} {\enquote {\bibinfo {title} {Low energy
  electron diffraction studies of gas adsorption on the platinum (100) single
  crystal surface},}\ }\href {\doibase 10.1016/0039-6028(68)90089-7} {\bibfield
   {journal} {\bibinfo  {journal} {Surface Science}\ }\textbf {\bibinfo
  {volume} {12}},\ \bibinfo {pages} {405} (\bibinfo {year} {1968})}\BibitemShut
  {NoStop}%
\bibitem [{\citenamefont {Pietzsch}\ \emph {et~al.}(2000)\citenamefont
  {Pietzsch}, \citenamefont {Kubetzka}, \citenamefont {Bode},\ and\
  \citenamefont {Wiesendanger}}]{Pietzsch2000}%
  \BibitemOpen
  \bibfield  {author} {\bibinfo {author} {\bibfnamefont {O.}~\bibnamefont
  {Pietzsch}}, \bibinfo {author} {\bibfnamefont {A.}~\bibnamefont {Kubetzka}},
  \bibinfo {author} {\bibfnamefont {M.}~\bibnamefont {Bode}}, \ and\ \bibinfo
  {author} {\bibfnamefont {R.}~\bibnamefont {Wiesendanger}},\ }\bibfield
  {title} {\enquote {\bibinfo {title} {Real-space observation of dipolar
  antiferromagnetism in magnetic nanowires by spin-polarized scanning tunneling
  spectroscopy},}\ }\href {\doibase 10.1103/PhysRevLett.84.5212} {\bibfield
  {journal} {\bibinfo  {journal} {Phys. Rev. Lett.}\ }\textbf {\bibinfo
  {volume} {84}},\ \bibinfo {pages} {5212} (\bibinfo {year}
  {2000})}\BibitemShut {NoStop}%
\bibitem [{\citenamefont {Bode}\ \emph {et~al.}(2001)\citenamefont {Bode},
  \citenamefont {Pietzsch}, \citenamefont {Kubetzka}, \citenamefont {Heinze},\
  and\ \citenamefont {Wiesendanger}}]{Bode2001}%
  \BibitemOpen
  \bibfield  {author} {\bibinfo {author} {\bibfnamefont {M.}~\bibnamefont
  {Bode}}, \bibinfo {author} {\bibfnamefont {O.}~\bibnamefont {Pietzsch}},
  \bibinfo {author} {\bibfnamefont {A.}~\bibnamefont {Kubetzka}}, \bibinfo
  {author} {\bibfnamefont {S.}~\bibnamefont {Heinze}}, \ and\ \bibinfo {author}
  {\bibfnamefont {R.}~\bibnamefont {Wiesendanger}},\ }\bibfield  {title}
  {\enquote {\bibinfo {title} {Experimental evidence for intra-atomic
  noncollinear magnetism at thin film probe tips},}\ }\href {\doibase
  10.1103/PhysRevLett.86.2142} {\bibfield  {journal} {\bibinfo  {journal}
  {Phys. Rev. Lett.}\ }\textbf {\bibinfo {volume} {86}},\ \bibinfo {pages}
  {2142--2145} (\bibinfo {year} {2001})}\BibitemShut {NoStop}%
\bibitem [{\citenamefont {Meckler}\ \emph {et~al.}(2009)\citenamefont
  {Meckler}, \citenamefont {Mikuszeit}, \citenamefont {Pre\ss{}ler},
  \citenamefont {Vedmedenko}, \citenamefont {Pietzsch},\ and\ \citenamefont
  {Wiesendanger}}]{Meckler2009}%
  \BibitemOpen
  \bibfield  {author} {\bibinfo {author} {\bibfnamefont {S.}~\bibnamefont
  {Meckler}}, \bibinfo {author} {\bibfnamefont {N.}~\bibnamefont {Mikuszeit}},
  \bibinfo {author} {\bibfnamefont {A.}~\bibnamefont {Pre\ss{}ler}}, \bibinfo
  {author} {\bibfnamefont {E.~Y.}\ \bibnamefont {Vedmedenko}}, \bibinfo
  {author} {\bibfnamefont {O.}~\bibnamefont {Pietzsch}}, \ and\ \bibinfo
  {author} {\bibfnamefont {R.}~\bibnamefont {Wiesendanger}},\ }\bibfield
  {title} {\enquote {\bibinfo {title} {Real-space observation of a
  right-rotating inhomogeneous cycloidal spin spiral by spin-polarized scanning
  tunneling microscopy in a triple axes vector magnet},}\ }\href {\doibase
  10.1103/PhysRevLett.103.157201} {\bibfield  {journal} {\bibinfo  {journal}
  {Phys. Rev. Lett.}\ }\textbf {\bibinfo {volume} {103}},\ \bibinfo {pages}
  {157201} (\bibinfo {year} {2009})}\BibitemShut {NoStop}%
\bibitem [{\citenamefont {Bode}\ \emph {et~al.}(1998)\citenamefont {Bode},
  \citenamefont {Getzlaff},\ and\ \citenamefont
  {Wiesendanger}}]{PhysRevLett.81.4256}%
  \BibitemOpen
  \bibfield  {author} {\bibinfo {author} {\bibfnamefont {M.}~\bibnamefont
  {Bode}}, \bibinfo {author} {\bibfnamefont {M.}~\bibnamefont {Getzlaff}}, \
  and\ \bibinfo {author} {\bibfnamefont {R.}~\bibnamefont {Wiesendanger}},\
  }\bibfield  {title} {\enquote {\bibinfo {title} {Spin-polarized vacuum
  tunneling into the exchange-split surface state of Gd(0001)},}\ }\href
  {\doibase 10.1103/PhysRevLett.81.4256} {\bibfield  {journal} {\bibinfo
  {journal} {Phys. Rev. Lett.}\ }\textbf {\bibinfo {volume} {81}},\ \bibinfo
  {pages} {4256--4259} (\bibinfo {year} {1998})}\BibitemShut {NoStop}%
\bibitem [{\citenamefont {Berbil-Bautista}\ \emph {et~al.}(2007)\citenamefont
  {Berbil-Bautista}, \citenamefont {Krause}, \citenamefont {Bode},\ and\
  \citenamefont {Wiesendanger}}]{PhysRevB.76.064411}%
  \BibitemOpen
  \bibfield  {author} {\bibinfo {author} {\bibfnamefont {L.}~\bibnamefont
  {Berbil-Bautista}}, \bibinfo {author} {\bibfnamefont {S.}~\bibnamefont
  {Krause}}, \bibinfo {author} {\bibfnamefont {M.}~\bibnamefont {Bode}}, \ and\
  \bibinfo {author} {\bibfnamefont {R.}~\bibnamefont {Wiesendanger}},\
  }\bibfield  {title} {\enquote {\bibinfo {title} {Spin-polarized scanning
  tunneling microscopy and spectroscopy of ferromagnetic Dy(0001)/W(110)
  films},}\ }\href {\doibase 10.1103/PhysRevB.76.064411} {\bibfield  {journal}
  {\bibinfo  {journal} {Phys. Rev. B}\ }\textbf {\bibinfo {volume} {76}},\
  \bibinfo {pages} {064411} (\bibinfo {year} {2007})}\BibitemShut {NoStop}%
\bibitem [{\citenamefont {Ravli\ifmmode~\acute{c}\else \'{c}\fi{}}\ \emph
  {et~al.}(2003)\citenamefont {Ravli\ifmmode~\acute{c}\else \'{c}\fi{}},
  \citenamefont {Bode}, \citenamefont {Kubetzka},\ and\ \citenamefont
  {Wiesendanger}}]{PhysRevB.67.174411}%
  \BibitemOpen
  \bibfield  {author} {\bibinfo {author} {\bibfnamefont {R.}~\bibnamefont
  {Ravli\ifmmode~\acute{c}\else \'{c}\fi{}}}, \bibinfo {author} {\bibfnamefont
  {M.}~\bibnamefont {Bode}}, \bibinfo {author} {\bibfnamefont {A.}~\bibnamefont
  {Kubetzka}}, \ and\ \bibinfo {author} {\bibfnamefont {R.}~\bibnamefont
  {Wiesendanger}},\ }\bibfield  {title} {\enquote {\bibinfo {title}
  {Correlation of dislocation and domain structure of Cr(001) investigated by
  spin-polarized scanning tunneling microscopy},}\ }\href {\doibase
  10.1103/PhysRevB.67.174411} {\bibfield  {journal} {\bibinfo  {journal} {Phys.
  Rev. B}\ }\textbf {\bibinfo {volume} {67}},\ \bibinfo {pages} {174411}
  (\bibinfo {year} {2003})}\BibitemShut {NoStop}%
\bibitem [{\citenamefont {Kurz}\ \emph {et~al.}(2002)\citenamefont {Kurz},
  \citenamefont {Bihlmayer},\ and\ \citenamefont {Bl{\"u}gel}}]{Kurz_2002}%
  \BibitemOpen
  \bibfield  {author} {\bibinfo {author} {\bibfnamefont {Ph.}\ \bibnamefont
  {Kurz}}, \bibinfo {author} {\bibfnamefont {G.}~\bibnamefont {Bihlmayer}}, \
  and\ \bibinfo {author} {\bibfnamefont {S.}~\bibnamefont {Bl{\"u}gel}},\
  }\bibfield  {title} {\enquote {\bibinfo {title} {Magnetism and electronic
  structure of hcp Gd and the Gd(0001) surface},}\ }\href {\doibase
  10.1088/0953-8984/14/25/305} {\bibfield  {journal} {\bibinfo  {journal}
  {Journal of Physics: Condensed Matter}\ }\textbf {\bibinfo {volume} {14}},\
  \bibinfo {pages} {6353--6371} (\bibinfo {year} {2002})}\BibitemShut {NoStop}%
\bibitem [{\citenamefont {Arblaster}(2010)}]{Arblaster2010}%
  \BibitemOpen
  \bibfield  {author} {\bibinfo {author} {\bibfnamefont {J.~W.}\ \bibnamefont
  {Arblaster}},\ }\bibfield  {title} {\enquote {\bibinfo {title}
  {Crystallographic properties of iridium},}\ }\href {\doibase
  10.1595/147106710X493124} {\bibfield  {journal} {\bibinfo  {journal}
  {Platinum Metals Review}\ }\textbf {\bibinfo {volume} {54}},\ \bibinfo
  {pages} {93--102} (\bibinfo {year} {2010})}\BibitemShut {NoStop}%
\bibitem [{Sup()}]{Sup}%
  \BibitemOpen
  \href@noop {} {}\bibinfo {note} {See supplemental material [url]
  for detailed information regarding the preparation of clean, 
  oxygen-reconstructed, and TMO-covered fcc(001) surfaces, 
  the characterization of magnetic probe tips, 
  and SP-STM investigations of TMOs not discussed in the main text, 
  including Refs. [42-50].}\BibitemShut {Stop}%
\bibitem [{\citenamefont {Anderson}(1950)}]{Anderson_SuperExchange}%
  \BibitemOpen
  \bibfield  {author} {\bibinfo {author} {\bibfnamefont {P.~W.}\ \bibnamefont
  {Anderson}},\ }\bibfield  {title} {\enquote {\bibinfo {title}
  {Antiferromagnetism. theory of superexchange interaction},}\ }\href {\doibase
  10.1103/PhysRev.79.350} {\bibfield  {journal} {\bibinfo  {journal} {Phys.
  Rev.}\ }\textbf {\bibinfo {volume} {79}},\ \bibinfo {pages} {350--356}
  (\bibinfo {year} {1950})}\BibitemShut {NoStop}%
\bibitem [{\citenamefont {Fert}\ and\ \citenamefont {Levy}(1980)}]{FL1980}%
  \BibitemOpen
  \bibfield  {author} {\bibinfo {author} {\bibfnamefont {A.}~\bibnamefont
  {Fert}}\ and\ \bibinfo {author} {\bibfnamefont {Peter~M.}\ \bibnamefont
  {Levy}},\ }\bibfield  {title} {\enquote {\bibinfo {title} {Role of
  anisotropic exchange interactions in determining the properties of
  spin-glasses},}\ }\href {\doibase 10.1103/PhysRevLett.44.1538} {\bibfield
  {journal} {\bibinfo  {journal} {Phys. Rev. Lett.}\ }\textbf {\bibinfo
  {volume} {44}},\ \bibinfo {pages} {1538--1541} (\bibinfo {year}
  {1980})}\BibitemShut {NoStop}%
\bibitem [{\citenamefont {Dup{\'e}}\ \emph {et~al.}(2015)\citenamefont
  {Dup{\'e}}, \citenamefont {Bickel}, \citenamefont {Mokrousov}, \citenamefont
  {Otte}, \citenamefont {von Bergmann}, \citenamefont {Kubetzka}, \citenamefont
  {Heinze},\ and\ \citenamefont {Wiesendanger}}]{2015Dupe}%
  \BibitemOpen
  \bibfield  {author} {\bibinfo {author} {\bibfnamefont {B.}~\bibnamefont
  {Dup{\'e}}}, \bibinfo {author} {\bibfnamefont {J.E.}\ \bibnamefont {Bickel}},
  \bibinfo {author} {\bibfnamefont {Y.}~\bibnamefont {Mokrousov}}, \bibinfo
  {author} {\bibfnamefont {F.}~\bibnamefont {Otte}}, \bibinfo {author}
  {\bibfnamefont {K.}~\bibnamefont {von Bergmann}}, \bibinfo {author}
  {\bibfnamefont {A.}~\bibnamefont {Kubetzka}}, \bibinfo {author}
  {\bibfnamefont {S.}~\bibnamefont {Heinze}}, \ and\ \bibinfo {author}
  {\bibfnamefont {R.}~\bibnamefont {Wiesendanger}},\ }\bibfield  {title}
  {\enquote {\bibinfo {title} {Giant magnetization canting due to symmetry
  breaking in zigzag {Co} chains on {Ir(001)}},}\ }\href
  {http://stacks.iop.org/1367-2630/17/i=2/a=023014} {\bibfield  {journal}
  {\bibinfo  {journal} {New Journal of Physics}\ }\textbf {\bibinfo {volume}
  {17}},\ \bibinfo {pages} {023014} (\bibinfo {year} {2015})}\BibitemShut
  {NoStop}%
\bibitem [{\citenamefont {Bouaziz}\ \emph {et~al.}(2017)\citenamefont
  {Bouaziz}, \citenamefont {dos Santos~Dias}, \citenamefont {Ziane},
  \citenamefont {Benakki}, \citenamefont {Bl{\"u}gel},\ and\ \citenamefont
  {Lounis}}]{Bouaziz2017}%
  \BibitemOpen
  \bibfield  {author} {\bibinfo {author} {\bibfnamefont {J.}~\bibnamefont
  {Bouaziz}}, \bibinfo {author} {\bibfnamefont {M.}~\bibnamefont {dos
  Santos~Dias}}, \bibinfo {author} {\bibfnamefont {A.}~\bibnamefont {Ziane}},
  \bibinfo {author} {\bibfnamefont {M.}~\bibnamefont {Benakki}}, \bibinfo
  {author} {\bibfnamefont {S.}~\bibnamefont {Bl{\"u}gel}}, \ and\ \bibinfo
  {author} {\bibfnamefont {S.}~\bibnamefont {Lounis}},\ }\bibfield  {title}
  {\enquote {\bibinfo {title} {Chiral magnetism of magnetic adatoms generated
  by {Rashba} electrons},}\ }\href
  {http://stacks.iop.org/1367-2630/19/i=2/a=023010} {\bibfield  {journal}
  {\bibinfo  {journal} {New Journal of Physics}\ }\textbf {\bibinfo {volume}
  {19}},\ \bibinfo {pages} {023010} (\bibinfo {year} {2017})}\BibitemShut
  {NoStop}%
\bibitem [{\citenamefont {Grigoriev}\ \emph {et~al.}(2008)\citenamefont
  {Grigoriev}, \citenamefont {Chetverikov}, \citenamefont {Lott},\ and\
  \citenamefont {Schreyer}}]{GCY2008}%
  \BibitemOpen
  \bibfield  {author} {\bibinfo {author} {\bibfnamefont {S.~V.}\ \bibnamefont
  {Grigoriev}}, \bibinfo {author} {\bibfnamefont {Yu.~O.}\ \bibnamefont
  {Chetverikov}}, \bibinfo {author} {\bibfnamefont {D.}~\bibnamefont {Lott}}, \
  and\ \bibinfo {author} {\bibfnamefont {A.}~\bibnamefont {Schreyer}},\
  }\bibfield  {title} {\enquote {\bibinfo {title} {Field induced chirality in
  the helix structure of $\mathrm{Dy}/\mathrm{Y}$ multilayer films and
  experimental evidence for {Dzyaloshinskii-Moriya} interaction on the
  interfaces},}\ }\href {\doibase 10.1103/PhysRevLett.100.197203} {\bibfield
  {journal} {\bibinfo  {journal} {Phys. Rev. Lett.}\ }\textbf {\bibinfo
  {volume} {100}},\ \bibinfo {pages} {197203} (\bibinfo {year}
  {2008})}\BibitemShut {NoStop}%
\bibitem [{\citenamefont {Hammer}\ \emph
  {et~al.}(2016{\natexlab{b}})\citenamefont {Hammer}, \citenamefont {Meinel},
  \citenamefont {Krahn},\ and\ \citenamefont {Widdra}}]{PhysRevB.94.195406}%
  \BibitemOpen
  \bibfield  {author} {\bibinfo {author} {\bibfnamefont {R.}~\bibnamefont
  {Hammer}}, \bibinfo {author} {\bibfnamefont {K.}~\bibnamefont {Meinel}},
  \bibinfo {author} {\bibfnamefont {O.}~\bibnamefont {Krahn}}, \ and\ \bibinfo
  {author} {\bibfnamefont {W.}~\bibnamefont {Widdra}},\ }\bibfield  {title}
  {\enquote {\bibinfo {title} {Surface reconstruction of pt(001) quantitatively
  revisited},}\ }\href {\doibase 10.1103/PhysRevB.94.195406} {\bibfield
  {journal} {\bibinfo  {journal} {Phys. Rev. B}\ }\textbf {\bibinfo {volume}
  {94}},\ \bibinfo {pages} {195406} (\bibinfo {year}
  {2016}{\natexlab{b}})}\BibitemShut {NoStop}%
\bibitem [{\citenamefont {van Houselt}\ \emph {et~al.}(2008)\citenamefont {van
  Houselt}, \citenamefont {Gnielka}, \citenamefont {de~Brugh}, \citenamefont
  {Oncel}, \citenamefont {Kockmann}, \citenamefont {Heid}, \citenamefont
  {Bohnen}, \citenamefont {Poelsema},\ and\ \citenamefont
  {Zandvliet}}]{Houselt2008}%
  \BibitemOpen
  \bibfield  {author} {\bibinfo {author} {\bibfnamefont {A.}~\bibnamefont {van
  Houselt}}, \bibinfo {author} {\bibfnamefont {T.}~\bibnamefont {Gnielka}},
  \bibinfo {author} {\bibfnamefont {J.~M. J.~Aan}\ \bibnamefont {de~Brugh}},
  \bibinfo {author} {\bibfnamefont {N.}~\bibnamefont {Oncel}}, \bibinfo
  {author} {\bibfnamefont {D.}~\bibnamefont {Kockmann}}, \bibinfo {author}
  {\bibfnamefont {R.}~\bibnamefont {Heid}}, \bibinfo {author} {\bibfnamefont
  {K.-P.}\ \bibnamefont {Bohnen}}, \bibinfo {author} {\bibfnamefont
  {B.}~\bibnamefont {Poelsema}}, \ and\ \bibinfo {author} {\bibfnamefont
  {H.~J.~W.}\ \bibnamefont {Zandvliet}},\ }\bibfield  {title} {\enquote
  {\bibinfo {title} {Peierls instability in {P}t chains on {G}e(001)},}\ }\href
  {\doibase 10.1016/j.susc.2008.01.040} {\bibfield  {journal} {\bibinfo
  {journal} {Surface Science}\ }\textbf {\bibinfo {volume} {602}},\ \bibinfo
  {pages} {1731} (\bibinfo {year} {2008})}\BibitemShut {NoStop}%
\bibitem [{\citenamefont {Gilarowski}\ \emph {et~al.}(2000)\citenamefont
  {Gilarowski}, \citenamefont {M{\'{e}}ndez},\ and\ \citenamefont
  {Niehus}}]{Gilarowski2000}%
  \BibitemOpen
  \bibfield  {author} {\bibinfo {author} {\bibfnamefont {Gerhard}\ \bibnamefont
  {Gilarowski}}, \bibinfo {author} {\bibfnamefont {Javier}\ \bibnamefont
  {M{\'{e}}ndez}}, \ and\ \bibinfo {author} {\bibfnamefont {Horst}\
  \bibnamefont {Niehus}},\ }\bibfield  {title} {\enquote {\bibinfo {title}
  {Initial growth of cu on ir(100)-$(5\times1)$},}\ }\href {\doibase
  https://doi.org/10.1016/S0039-6028(99)01255-8} {\bibfield  {journal}
  {\bibinfo  {journal} {Surface Science}\ }\textbf {\bibinfo {volume} {448}},\
  \bibinfo {pages} {290} (\bibinfo {year} {2000})}\BibitemShut {NoStop}%
\bibitem [{\citenamefont {Moriya}(1960)}]{Mor1960}%
  \BibitemOpen
  \bibfield  {author} {\bibinfo {author} {\bibfnamefont {T.}~\bibnamefont
  {Moriya}},\ }\bibfield  {title} {\enquote {\bibinfo {title} {Anisotropic
  superexchange interaction and weak ferromagnetism},}\ }\href {\doibase
  10.1103/PhysRev.120.91} {\bibfield  {journal} {\bibinfo  {journal} {Phys.
  Rev.}\ }\textbf {\bibinfo {volume} {120}},\ \bibinfo {pages} {91--98}
  (\bibinfo {year} {1960})}\BibitemShut {NoStop}%
\bibitem [{\citenamefont {Moritz}\ \emph {et~al.}(2004)\citenamefont {Moritz},
  \citenamefont {Garcia}, \citenamefont {Toussaint}, \citenamefont {Dieny},\
  and\ \citenamefont {Nozi{\`e}res}}]{Mor2004}%
  \BibitemOpen
  \bibfield  {author} {\bibinfo {author} {\bibfnamefont {J.}~\bibnamefont
  {Moritz}}, \bibinfo {author} {\bibfnamefont {F.}~\bibnamefont {Garcia}},
  \bibinfo {author} {\bibfnamefont {J.~C.}\ \bibnamefont {Toussaint}}, \bibinfo
  {author} {\bibfnamefont {B.}~\bibnamefont {Dieny}}, \ and\ \bibinfo {author}
  {\bibfnamefont {J.~P.}\ \bibnamefont {Nozi{\`e}res}},\ }\bibfield  {title}
  {\enquote {\bibinfo {title} {Orange peel coupling in multilayers with
  perpendicular magnetic anisotropy: Application to (co/pt)-based
  exchange-biased spin-valves},}\ }\href
  {https://doi.org/10.1209/epl/i2003-10063-9} {\bibfield  {journal} {\bibinfo
  {journal} {Europhys. Lett.}\ }\textbf {\bibinfo {volume} {65}},\ \bibinfo
  {pages} {123--129} (\bibinfo {year} {2004})}\BibitemShut {NoStop}%
\bibitem [{\citenamefont {Stoeffler}(2004)}]{Stoeffler2004}%
  \BibitemOpen
  \bibfield  {author} {\bibinfo {author} {\bibfnamefont {D.}~\bibnamefont
  {Stoeffler}},\ }\bibfield  {title} {\enquote {\bibinfo {title} {Ab initio
  study of the fe intra- and inter-layer magnetic order in fe/ir(001)
  superlattices},}\ }\href {\doibase 10.1140/epjb/e2004-00061-9} {\bibfield
  {journal} {\bibinfo  {journal} {The European Physical Journal B - Condensed
  Matter and Complex Systems}\ }\textbf {\bibinfo {volume} {37}},\ \bibinfo
  {pages} {311--320} (\bibinfo {year} {2004})}\BibitemShut {NoStop}%
\bibitem [{\citenamefont {Johnson}\ \emph {et~al.}(2000)\citenamefont
  {Johnson}, \citenamefont {Ge}, \citenamefont {Titmuss},\ and\ \citenamefont
  {King}}]{Johnson2000}%
  \BibitemOpen
  \bibfield  {author} {\bibinfo {author} {\bibfnamefont {K.}~\bibnamefont
  {Johnson}}, \bibinfo {author} {\bibfnamefont {Q.}~\bibnamefont {Ge}},
  \bibinfo {author} {\bibfnamefont {S.}~\bibnamefont {Titmuss}}, \ and\
  \bibinfo {author} {\bibfnamefont {D.~A.}\ \bibnamefont {King}},\ }\bibfield
  {title} {\enquote {\bibinfo {title} {Unusual bridged site for adsorbed oxygen
  adatoms: Theory and experiment for {I}r\{100\}--$(1 \times 2)$-{O}},}\ }\href
  {https://aip.scitation.org/doi/10.1063/1.481709} {\bibfield  {journal}
  {\bibinfo  {journal} {The Journal of Chemical Physics}\ }\textbf {\bibinfo
  {volume} {112}},\ \bibinfo {pages} {10460--10466} (\bibinfo {year}
  {2000})}\BibitemShut {NoStop}%
\bibitem [{\citenamefont {Ferstl}\ \emph
  {et~al.}(2016{\natexlab{b}})\citenamefont {Ferstl}, \citenamefont {Schmitt},
  \citenamefont {Schneider}, \citenamefont {Hammer}, \citenamefont {Michl},\
  and\ \citenamefont {M\"uller}}]{Ferstl-OIr}%
  \BibitemOpen
  \bibfield  {author} {\bibinfo {author} {\bibfnamefont {P.}~\bibnamefont
  {Ferstl}}, \bibinfo {author} {\bibfnamefont {T.}~\bibnamefont {Schmitt}},
  \bibinfo {author} {\bibfnamefont {M.~A.}\ \bibnamefont {Schneider}}, \bibinfo
  {author} {\bibfnamefont {L.}~\bibnamefont {Hammer}}, \bibinfo {author}
  {\bibfnamefont {A.}~\bibnamefont {Michl}}, \ and\ \bibinfo {author}
  {\bibfnamefont {S.}~\bibnamefont {M\"uller}},\ }\bibfield  {title} {\enquote
  {\bibinfo {title} {Structure and ordering of oxygen on unreconstructed
  {I}r(100)},}\ }\href {\doibase 10.1103/PhysRevB.93.235406} {\bibfield
  {journal} {\bibinfo  {journal} {Phys. Rev. B}\ }\textbf {\bibinfo {volume}
  {93}},\ \bibinfo {pages} {235406} (\bibinfo {year}
  {2016}{\natexlab{b}})}\BibitemShut {NoStop}%
\bibitem [{\citenamefont {Elmers}\ \emph {et~al.}(1999)\citenamefont {Elmers},
  \citenamefont {Hauschild},\ and\ \citenamefont {Gradmann}}]{Elmers1999}%
  \BibitemOpen
  \bibfield  {author} {\bibinfo {author} {\bibfnamefont {H.~J.}\ \bibnamefont
  {Elmers}}, \bibinfo {author} {\bibfnamefont {J.}~\bibnamefont {Hauschild}}, \
  and\ \bibinfo {author} {\bibfnamefont {U.}~\bibnamefont {Gradmann}},\
  }\bibfield  {title} {\enquote {\bibinfo {title} {Onset of perpendicular
  magnetization in nanostripe arrays of {F}e on stepped {W(110)} surfaces},}\
  }\href {\doibase 10.1103/PhysRevB.59.3688} {\bibfield  {journal} {\bibinfo
  {journal} {Phys. Rev. B}\ }\textbf {\bibinfo {volume} {59}},\ \bibinfo
  {pages} {3688--3695} (\bibinfo {year} {1999})}\BibitemShut {NoStop}%
\bibitem [{\citenamefont {Bode}\ \emph {et~al.}(2002)\citenamefont {Bode},
  \citenamefont {Heinze}, \citenamefont {Kubetzka}, \citenamefont {Pietzsch},
  \citenamefont {Nie}, \citenamefont {Bihlmayer}, \citenamefont {Bl\"ugel},\
  and\ \citenamefont {Wiesendanger}}]{Bode2002}%
  \BibitemOpen
  \bibfield  {author} {\bibinfo {author} {\bibfnamefont {M.}~\bibnamefont
  {Bode}}, \bibinfo {author} {\bibfnamefont {S.}~\bibnamefont {Heinze}},
  \bibinfo {author} {\bibfnamefont {A.}~\bibnamefont {Kubetzka}}, \bibinfo
  {author} {\bibfnamefont {O.}~\bibnamefont {Pietzsch}}, \bibinfo {author}
  {\bibfnamefont {X.}~\bibnamefont {Nie}}, \bibinfo {author} {\bibfnamefont
  {G.}~\bibnamefont {Bihlmayer}}, \bibinfo {author} {\bibfnamefont
  {S.}~\bibnamefont {Bl\"ugel}}, \ and\ \bibinfo {author} {\bibfnamefont
  {R.}~\bibnamefont {Wiesendanger}},\ }\bibfield  {title} {\enquote {\bibinfo
  {title} {Magnetization-direction-dependent local electronic structure probed
  by scanning tunneling spectroscopy},}\ }\href {\doibase
  10.1103/PhysRevLett.89.237205} {\bibfield  {journal} {\bibinfo  {journal}
  {Phys. Rev. Lett.}\ }\textbf {\bibinfo {volume} {89}},\ \bibinfo {pages}
  {237205} (\bibinfo {year} {2002})}\BibitemShut {NoStop}%
\bibitem [{\citenamefont {Pietzsch}\ \emph {et~al.}(2001)\citenamefont
  {Pietzsch}, \citenamefont {Kubetzka}, \citenamefont {Bode},\ and\
  \citenamefont {Wiesendanger}}]{Pietzsch2001}%
  \BibitemOpen
  \bibfield  {author} {\bibinfo {author} {\bibfnamefont {O.}~\bibnamefont
  {Pietzsch}}, \bibinfo {author} {\bibfnamefont {A.}~\bibnamefont {Kubetzka}},
  \bibinfo {author} {\bibfnamefont {M.}~\bibnamefont {Bode}}, \ and\ \bibinfo
  {author} {\bibfnamefont {R.}~\bibnamefont {Wiesendanger}},\ }\bibfield
  {title} {\enquote {\bibinfo {title} {Observation of magnetic hysteresis at
  the nanometer scale by spin-polarized scanning tunneling spectroscopy},}\
  }\href {\doibase 10.1126/science.1060513} {\bibfield  {journal} {\bibinfo
  {journal} {Science}\ }\textbf {\bibinfo {volume} {292}},\ \bibinfo {pages}
  {2053--2056} (\bibinfo {year} {2001})}\BibitemShut {NoStop}%
\bibitem [{\citenamefont {Kubetzka}\ \emph {et~al.}(2003)\citenamefont
  {Kubetzka}, \citenamefont {Pietzsch}, \citenamefont {Bode},\ and\
  \citenamefont {Wiesendanger}}]{Kubetzka2003}%
  \BibitemOpen
  \bibfield  {author} {\bibinfo {author} {\bibfnamefont {A.}~\bibnamefont
  {Kubetzka}}, \bibinfo {author} {\bibfnamefont {O.}~\bibnamefont {Pietzsch}},
  \bibinfo {author} {\bibfnamefont {M.}~\bibnamefont {Bode}}, \ and\ \bibinfo
  {author} {\bibfnamefont {R.}~\bibnamefont {Wiesendanger}},\ }\bibfield
  {title} {\enquote {\bibinfo {title} {Spin-polarized scanning tunneling
  microscopy study of $360^{\circ}$ walls in an external magnetic field},}\
  }\href {\doibase 10.1103/PhysRevB.67.020401} {\bibfield  {journal} {\bibinfo
  {journal} {Phys. Rev. B}\ }\textbf {\bibinfo {volume} {67}},\ \bibinfo
  {pages} {020401} (\bibinfo {year} {2003})}\BibitemShut {NoStop}%
\bibitem [{\citenamefont {Bode}\ \emph {et~al.}(2003)\citenamefont {Bode},
  \citenamefont {Kubetzka}, \citenamefont {Heinze}, \citenamefont {Pietzsch},
  \citenamefont {Wiesendanger}, \citenamefont {Heide}, \citenamefont {Nie},
  \citenamefont {Bihlmayer},\ and\ \citenamefont {gel}}]{Bode2003}%
  \BibitemOpen
  \bibfield  {author} {\bibinfo {author} {\bibfnamefont {M.}~\bibnamefont
  {Bode}}, \bibinfo {author} {\bibfnamefont {A.}~\bibnamefont {Kubetzka}},
  \bibinfo {author} {\bibfnamefont {S.}~\bibnamefont {Heinze}}, \bibinfo
  {author} {\bibfnamefont {O.}~\bibnamefont {Pietzsch}}, \bibinfo {author}
  {\bibfnamefont {R.}~\bibnamefont {Wiesendanger}}, \bibinfo {author}
  {\bibfnamefont {M.}~\bibnamefont {Heide}}, \bibinfo {author} {\bibfnamefont
  {X.}~\bibnamefont {Nie}}, \bibinfo {author} {\bibfnamefont {G.}~\bibnamefont
  {Bihlmayer}}, \ and\ \bibinfo {author} {\bibfnamefont {S.~Bl{\"u}gel}},\ }\bibfield  
  {title} {\enquote {\bibinfo {title} {Spin-orbit induced
  local band structure variations revealed by scanning tunnelling
  spectroscopy},}\ }\href {\doibase 10.1088/0953-8984/15/5/320} {\bibfield
  {journal} {\bibinfo  {journal} {Journal of Physics: Condensed Matter}\
  }\textbf {\bibinfo {volume} {15}},\ \bibinfo {pages} {S679--S692} (\bibinfo
  {year} {2003})}\BibitemShut {NoStop}%
\bibitem [{\citenamefont {Vedmedenko}\ \emph {et~al.}(2004)\citenamefont
  {Vedmedenko}, \citenamefont {Kubetzka}, \citenamefont {von Bergmann},
  \citenamefont {Pietzsch}, \citenamefont {Bode}, \citenamefont {Kirschner},
  \citenamefont {Oepen},\ and\ \citenamefont {Wiesendanger}}]{Vedmedenko2004}%
  \BibitemOpen
  \bibfield  {author} {\bibinfo {author} {\bibfnamefont {E.~Y.}\ \bibnamefont
  {Vedmedenko}}, \bibinfo {author} {\bibfnamefont {A.}~\bibnamefont
  {Kubetzka}}, \bibinfo {author} {\bibfnamefont {K.}~\bibnamefont {von
  Bergmann}}, \bibinfo {author} {\bibfnamefont {O.}~\bibnamefont {Pietzsch}},
  \bibinfo {author} {\bibfnamefont {M.}~\bibnamefont {Bode}}, \bibinfo {author}
  {\bibfnamefont {J.}~\bibnamefont {Kirschner}}, \bibinfo {author}
  {\bibfnamefont {H.~P.}\ \bibnamefont {Oepen}}, \ and\ \bibinfo {author}
  {\bibfnamefont {R.}~\bibnamefont {Wiesendanger}},\ }\bibfield  {title}
  {\enquote {\bibinfo {title} {Domain wall orientation in magnetic
  nanowires},}\ }\href {\doibase 10.1103/PhysRevLett.92.077207} {\bibfield
  {journal} {\bibinfo  {journal} {Phys. Rev. Lett.}\ }\textbf {\bibinfo
  {volume} {92}},\ \bibinfo {pages} {077207} (\bibinfo {year}
  {2004})}\BibitemShut {NoStop}%
\bibitem [{\citenamefont {Elmers}\ \emph {et~al.}(1995)\citenamefont {Elmers},
  \citenamefont {Hauschild}, \citenamefont {Fritzsche}, \citenamefont {Liu},
  \citenamefont {Gradmann},\ and\ \citenamefont {K\"ohler}}]{Elmers1995}%
  \BibitemOpen
  \bibfield  {author} {\bibinfo {author} {\bibfnamefont {H.~J.}\ \bibnamefont
  {Elmers}}, \bibinfo {author} {\bibfnamefont {J.}~\bibnamefont {Hauschild}},
  \bibinfo {author} {\bibfnamefont {H.}~\bibnamefont {Fritzsche}}, \bibinfo
  {author} {\bibfnamefont {G.}~\bibnamefont {Liu}}, \bibinfo {author}
  {\bibfnamefont {U.}~\bibnamefont {Gradmann}}, \ and\ \bibinfo {author}
  {\bibfnamefont {U.}~\bibnamefont {K\"ohler}},\ }\bibfield  {title} {\enquote
  {\bibinfo {title} {Magnetic frustration in ultrathin {F}e films},}\ }\href
  {\doibase 10.1103/PhysRevLett.75.2031} {\bibfield  {journal} {\bibinfo
  {journal} {Phys. Rev. Lett.}\ }\textbf {\bibinfo {volume} {75}},\ \bibinfo
  {pages} {2031} (\bibinfo {year} {1995})}\BibitemShut {NoStop}%
\end{thebibliography}
\end{document}